\documentclass[oldversion]{aa}

\usepackage{natbib}
\usepackage{epsfig}
\usepackage{txfonts}

\bibpunct{(}{)}{;}{a}{}{,} 

\begin{document}

\title{Nucleosynthesis-relevant conditions in neutrino-driven
              supernova outflows}
\subtitle{I.~Spherically symmetric hydrodynamic simulations}

\author{A. Arcones, H.-Th. Janka, and L. Scheck}

\offprints{} 
\mail{thj@mpa-garching.mpg.de}

\date{received; accepted}      
    
\institute{Max-Planck-Institut f\"ur Astrophysik,
           Karl-Schwarzschild-Stra{\ss}e 1, 
           D-85741 Garching, Germany}

\abstract{ We investigate the behavior and consequences of the reverse
shock that terminates the supersonic expansion of the baryonic
wind which is driven by neutrino heating off the surface of 
(non-magnetized) new-born neutron stars in supernova cores. To this 
end we perform long-time hydrodynamic simulations in spherical symmetry. 
In agreement with previous relativistic wind studies, 
we find that the neutrino-driven outflow
accelerates to supersonic velocities and in case of a compact, 
$\sim\,$1.4$\,M_\odot$ (gravitational mass) neutron star 
with a radius of about 10$\,$km, the wind reaches entropies of about 
100$\,k_{\mathrm B}$ per nucleon. The wind, however, is strongly
influenced by the environment of the supernova core. It
is decelerated and shock-heated abruptly by a termination
shock that forms when the supersonic outflow collides 
with the slower preceding supernova ejecta. 
The radial position of this reverse shock varies with time and depends
on the strength of the neutrino wind and the explosion conditions
in progenitor stars with different masses and structure.
Its basic properties and behavior can be understood by simple analytic
considerations. 
We demonstrate that the entropy of the matter going through the reverse 
shock can increase to a multiple of the asymptotic wind value.
Seconds after the onset of the explosion it therefore
can exceed 400$\,k_{\mathrm B}$ per nucleon in low-mass 
progenitors around 10$\,M_\odot$, where the supernova shock and the
reverse shock propagate outward quickly. The temperature of the 
shocked wind has typically dropped to about or less than
$10^9\,$K, and density and temperature in the shock-decelerated matter
continue to decrease only very slowly. For more massive progenitors
with bigger and denser metal cores, the explosion expands more slowly
so that the termination shock stays at smaller radii and affects the
wind at higher temperatures and densities. In this case the termination 
shock might play a non-negligible, strongly time- and progenitor-dependent
role in discussing supernova nucleosynthesis.
}
\authorrunning{Arcones et al.}
\titlerunning{Neutrino-driven supernova outflows}

\maketitle

\label{sec:introduction}
   
Neutron stars are born as extremely hot and dense remnants at the center of
exploding massive stars. Shortly after their formation they heat up to
temperatures that can reach tens of MeV and become even higher than 50$\,$MeV
for soft nuclear equations of state \cite[e.g.,][]{Burrows86, Keil95, Pons99}.
Their gravitational binding energy is carried away by neutrinos, which are
abundantly produced at such conditions. On a timescale of seconds these
neutrinos diffuse out of the interior of the star and escape from their mean
surface of last scattering, the neutrinosphere. By the associated energy and
lepton number loss, the hot, still proton-rich and neutrino-opaque
proto-neutron star thus evolves to the final cold, neutron-rich and
neutrino-transparent remnant during roughly the first minute of its life.

Outside of the neutrinosphere the radiated neutrinos, which have typical
energies of 10--20$\,$MeV, travel through a layer with a very steep density
gradient and decreasing temperature. It is unavoidable that the residual
interactions of the high-energy neutrinos with the cooler stellar matter
deposit energy in this region. This energy transfer does not allow the
``surface'' layers of the hot, neutrino-cooling neutron star to remain in
hydrostatic equilibrium, but leads to mass loss at a low rate in a
neutrino-driven outflow of baryonic matter \citep{Duncan86,Woosley92}. This
outflow, the so-called ``neutrino-driven wind'', unavoidably accompanies the
birth of a hot, neutrino-cooling neutron star, independent of the details of
the not finally understood supernova explosion mechanism \citep[in case of
sufficiently strong magnetic fields, however, there my be magnetically driven
outflow instead of a thermally-driven wind; see][]{Metzger06}.  The mass loss
of the nascent neutron star begins after the supernova explosion has been
launched and continues until the neutron star is essentially transparent to
neutrinos.  This flow of baryonic matter is a rapidly expanding and cooling
high-entropy environment, in which an $\alpha$-rich freeze-out can lead to the
production of elements heavier than the iron group.  For sufficiently large
neutron excess, the $\alpha$-rich freeze-out can merge smoothly into an
r-process \citep{Woosley92b,Meyer92}.

In fact, a number of parameters has been recognized to determine the
possibility of r-process nucleosynthesis in the neutrino-wind environment: The
neutron-to-proton ratio in the wind, expressed in terms of the
electron-to-baryon ratio or electron faction $Y_{\mathrm{e}}$; the expansion timescale,
$\tau$, which decides how fast the temperature and density of the outflowing
matter drop; and the wind entropy per nucleon, $s$, as a measure of the
photon-to-baryon ratio of the environment \citep{Witti94,Qian96,Hoffman96}. The
entropy of the wind is typically tens to more than 100 $k_{\mathrm{B}}$ per
nucleon, making the wind environment a candidate for the so-called high-entropy
r-process \citep{Meyer92,Meyer94}. In addition, the mass loss rate decides
whether the wind could be the major source of the observed galactic abundances
of r-process material. These wind parameters depend on the neutron star
properties, in particular on the gravitational field of the neutron star and
thus its mass and radius, and on the neutrino emission of the neutron star,
i.e., the time-dependent luminosities and spectra of the radiated neutrinos
\citep{Qian96}. Since it is mainly the absorption of electron neutrinos,
$\nu_{\mathrm{e}}$, and antineutrinos, $\bar\nu_{\mathrm{e}}$, on free neutrons and protons,
respectively, which heats the stellar gas and is responsible for driving the
mass loss and for setting the electron fraction in the ejected gas, the
emission properties of these neutrinos are most relevant (but see 
\citealt{Wanajo06}) for a suggestion that additional energy input to the wind by
$\nu\bar\nu$-annihilation in case of largely anisotropic neutrino emission from
the nascent neutron star could lead to a decisive increase of the wind
entropy).

Only in case the neutrino-driven outflow becomes supersonic beyond a critical
point, the sonic point, it truly deserves the name ``wind''. In such wind
solutions the physical conditions at the neutrinosphere and behind the
supernova shock are causally disconnected. The presence of the sonic point
unambiguously determines the solution for a given value of the driving
luminosity.  Wind solutions possess the highest (``critical'') mass loss rate
(and the lowest specific total energy of the ejected matter) for a given
neutrino luminosity. Physical solutions with larger mass loss rates (and lower
specific total energy) do not exist. Lower mass loss rates (higher specific
total energies) correspond to ``breeze solutions''
\citep{Takahashi97,Otsuki00}. In these, the outflow velocity reaches a maximum
and then decreases again to asymptote to zero at infinity. The whole region
between the proto-neutron star surface and the outer boundary of the considered
outflow is therefore in sonic contact. While wind solutions are characterized
by a continuously rising velocity and decreasing temperature, the temperature
of breezes level off to a constant value at large radii where the flow is
dominated by internal instead of kinetic energy. This limiting value of the
temperature at large distances from the neutron star is an additional
characteristic parameter of breeze solutions.

Transsonic neutrino-driven winds in the context of supernova explosions and
nucleosynthesis were investigated by means of hydrodynamic simulations
\citep{Woosley92}, analytic discussion \citep{Qian96,Cardall97}, and numerical
solutions of the steady-state wind equations \citep{Otsuki00,Thompson01}.
\citet{Otsuki00} discussed the difference between winds and breezes, but like
\citet{Wanajo01} they concentrated on the subsonic solutions for their
nucleosynthesis calculations, mainly because these allowed them to set a
boundary value of the temperature at some large radius. This was understood to
mimic the transition of the wind into a dense shell of ejecta behind the
outgoing supernova shock, the presence of which hampered the free expansion of
the wind. Such a behavior was found in calculations of supernova explosions by
the Livermore group, which were employed in the r-process studies of
\citet{Woosley92b}, \citet{Woosley94}, and \citet{Hoffman96}, and also in
hydrodynamic simulations of neutrino-driven outflows by \citet{Witti94}
and \citet{Takahashi94}, which were started from postbounce models
provided by the Livermore group.  The outflow trajectories in these simulations
showed temperature and density declining asymptotically to nearly constant
values, which were reached when the flow was gradually decelerated upon
catching up with the slower, earlier ejecta behind the supernova shock.
\citet{Sumiyoshi00} and \citet{Terasawa02} also referred to this
behavior for using an artificially imposed constant pressure at the outer
boundary in their Lagrangian hydrodynamic simulations of neutrino-driven mass
ejection.  The external pressure produced outflow deceleration similar to that
found in the previous supernova models.

Applying modern, high-resolution shock-capturing schemes and a better numerical
resolution to long-time hydrodynamic simulations of supernova explosions,
\citet{Janka95} and \citet{Burrows95} (see also the more recent models of
\citealt{Buras06a} used for nucleosynthesis studies by \citealt{Pruet05})
discovered the formation of a wind termination shock caused by the collision of
a transsonic neutrino-driven wind with the dense, slower ejecta shell behind
the supernova shock.  So far, however, this reverse shock, which leads to an
abrupt deceleration and shock heating of the fast wind, has not received much
attention. Subtle, potentially significant effects in the r-process
nucleosynthesis that may depend in interesting ways on the location of and
strength of the reverse shock were found by \citet{Thompson01}.  Although these
authors mentioned a rather modest reheating of the wind material by the reverse
shock passage (that causes a increase of the specific entropy of
10$\,k_{\mathrm{B}}$ per nucleon), they obtained a considerably enhanced
production of third-peak r-process nuclei due to a slower postshock expansion
and a significantly higher temperature (0.05$\,$MeV instead of 0.01$\,$MeV for
unshocked winds) at the time the r-process freeze-out happens. Also
\citet{Wanajo02}, alluding to the possibility of a wind termination shock,
introduced a freeze-out value $T_{\mathrm f}$ as the final temperature of the
wind, i.e., they limited the temperature (and density) decrease in the
supersonic wind by a chosen lower value.  The choice of this temperature was,
naturally, to some degree ad hoc, although \citet{Wanajo02} justified it by
nucleosynthesis considerations. A systematic and detailed exploration of the
formation of the wind termination shock, of its hydrodynamical effects on the
wind properties, and of its nucleosynthetic consequences, however, is still
lacking.

In this paper we will study the time-dependent evolution of the wind
termination shock in different progenitors with spherically symmetric (1D)
models. For this purpose we perform simulations of neutrino-driven explosions,
employing the approximations to the full supernova physics used in previous
works \citep{Scheck04,Scheck06}. The neutron star in our simulations is
replaced by a contracting inner boundary at which neutrino luminosities are
imposed such that supernova explosions with a typical explosion energy of
1--$2\times 10^{51}\,$erg$\,=\,$1--$2\,$bethe (B) are triggered by neutrino
heating. The subsequent explosion and evolution of the relic neutron star is
followed until 10$\,$s after core bounce. Varying the neutron star contraction,
which depends on the incompletely known high-density equation of state, and the
time-dependent neutrino emission from the forming neutron star, we will also
investigate the sensitivity of the reverse shock effects on the neutrino-wind
properties. Our results suggest that wind termination shocks are a robust,
long-lasting feature in the supernova core just like the outgoing supernova
shock and the neutrino-driven wind are. Of course, since a final understanding
of the explosion mechanism of core-collapse supernovae is still missing
\citep[see, e.g.][and references therein]{Buras03,Buras06a,Buras06b} and
because we excise the neutron star at the grid center instead of simulating its
neutrino-cooling evolution and, moreover, make severe approximations to the
important neutrino transport \citep[see Sect.~2.2 and Appendix~D
in][]{Scheck06}, our calculations will not be able to yield final answers.
Nevertheless, our results are suitable for discussing fundamental properties of
the wind termination shock and for developing a basic understanding of the role
of this so far not well studied aspect of supernova explosions.

Our paper is structured in the following way. In Sect.~\ref{sec:numerics} we
will briefly describe the numerical approach taken in this work.  In
Sect.~\ref{sec:relativity} we will present some comparisons we made with
previous work, in particular testing our approximative treatment of
relativistic effects. Section~\ref{sec:1Dresults} contains our results for
spherically symmetric simulations, providing a reference case and then
investigating varied conditions (neutron star radius, neutrino luminosities) at
the lower grid boundary and different progenitor stars.  In addition, we will
present an analytic discussion which allows one to basically understand the
behavior and the properties of the wind termination shock.
Section~\ref{sec:conclusions} will finish with a summary and 
conclusions.

\section{Numerical aspects}
\label{sec:numerics}

\subsection{Hydrodynamics and neutrino treatment}
\label{sec:hydro}

The simulations of this paper were carried out with the neutrino-hydrodynamics
code and the microphysics described by \citet{Scheck06}. The hydrodynamics
module is a version of the \textsc{Prometheus} code which is based on a direct
Eulerian implementation of the Piecewise Parabolic Method (PPM) of
\citet{Colella84}. It is a high-resolution shock capturing scheme and performs
a conservative, explicit integration of the Newtonian hydrodynamics equations
with third-order accuracy in space and second-order accuracy in time (see,
e.g., \citealt{Kifonidis03} and references therein).  General relativistic (GR)
gravity is approximated by using an ``effective relativistic potential'' (for
details, see Sect.~\ref{sec:gravity}). This approach accounts for the deeper
gravitational well in GR, but still works in Minkowski spacetime and ignores
the effects of relativistic kinematics. This was shown to yield excellent
agreement with full GR calculations during the pre-explosion phase after core
bounce \citep{Liebendorfer05,Marek06} and will be demonstrated to
also be a good approximation to full GR solutions of neutrino-driven winds in
Sect.~\ref{sec:relativity}.

The equation of state used in the simulations presented here is valid below
densities of roughly $10^{13}\,$g$\,$cm$^{-3}$ where non-ideal effects due to
strong interactions between nucleons can be safely ignored. It was used before
in the calculations by \citet{Janka96}, \cite{Kifonidis03}, and
\cite{Scheck06}.  Neutrons, protons, $\alpha$-particles and a representative
heavy nucleus of the iron group (chosen to be $^{54}\mathrm{Mn}$) are assumed
to be nonrelativistic Boltzmann-gases in nuclear statistical equilibrium.
Electrons and positrons are treated as Fermi-gases of arbitrary degeneracy and
arbitrary degree of relativity, and photon contributions are included as well.
Pressure and energy are corrected for Coulomb effects due to the
electromagnetic interactions between nucleons and the surrounding sea of
charged leptons.

The transport of neutrinos and antineutrinos of all flavors is based on a
computationally very efficient, analytic integration along characteristics of
the frequency-integrated zeroth-order moment equations of the Boltzmann
equation for neutrino number and energy (for details, see the Appendix of
\citealt{Scheck06}). The integration yields the neutrino number and energy
fluxes as functions of time and radius.  Our approach thus accounts for the
luminosity contributions due to the accretion on the forming neutron star. The
neutrino spectra are assumed to have Fermi-Dirac shape with a spectral
temperature that is determined from the ratio of neutrino-energy to
neutrino-number flux. Therefore in general the spectral temperature is
different from the local gas temperature. The closure of the neutrino number
and energy equations is achieved by employing the flux factor $f(r,t)=F/(Ec)$,
which couples the local energy (or number) flux with the neutrino energy (or
number) density.  For $f(r,t)$ we use a prescribed function which was
determined by fits to Monte Carlo transport results \citep{Janka91b}. This
yields a reasonably good approximation in the transparent and semi-transparent
regimes but is not designed to accurately reproduce the diffusion limit at very
high optical depths (where due to numerical reasons the applicability of the
approach is anyway strongly constrained by the need of very fine grid zoning).
The neutrino source terms in the transport equations and therefore the source
terms for lepton number, energy, and momentum in the hydrodynamics equations
include the most relevant neutrino-matter interactions \citep[cf.][]{Scheck06}.

We note that the steepening density gradient near the neutron star surface
requires extremely fine grid zoning for getting converged results of the
neutrino-driven outflow. We typically use about 1000 non-equidistant radial
mesh points.

\subsection{Boundary treatment}
\label{sec:boundary}

In our simulations we replace the inner core of the neutron star (usually
roughly 1$\,M_\odot$ of baryonic matter) by an inner Lagrangian boundary of our
grid, whose prescribed contraction is supposed to mimic the shrinking of the
nascent neutron star as it loses energy and lepton number by neutrino emission.
Using this inner boundary, which typically is located at a $\nu_{\mathrm{e}}$ optical
depth of more than 100 and a density of $\rho_{\mathrm{ib}} \ga
10^{13}\,$g$\,$cm$^{-3}$, does not only allow us to apply the simple neutrino
transport approximation described above, but also gives us the freedom to vary
the time-evolution of the neutron star radius and of the core neutrino fluxes
imposed at the inner grid boundary.  This makes sense because both the equation
of state of hot neutron star matter and the neutrino transport in nascent
neutron stars are not finally understood. Changing the inner boundary
conditions thus allows us to investigate the differences resulting from
different explosion energies and timescales and from a different evolution of
the neutrino-wind power in a given progenitor.

Three parameters serve us to describe the motion of the inner boundary:
$R_{\mathrm{i}}$, $R_{\mathrm{f}}$, and $t_0$. The initial radius
$R_{\mathrm{i}}$ is the radius of the inner core that we chose to excise from
the postbounce models we start our simulations from, $R_{\mathrm{f}}$ is the
final radius of this core for time $t \to \infty$, and $t_0$ is the timescale
of an exponential contraction according to the expression
\begin{equation}
R_{\mathrm{ib}}(t) = R_{\mathrm{f}}+(R_{\mathrm{i}}-R_{\mathrm{f}})
\mathrm{e}^{-t/t_{0}} \ .
\label{eq:ibmotion}
\end{equation}
Our standard choice of $t_0 = 0.1\,$s reproduces the contraction of the excised
core during the first few hundred milliseconds after bounce as found in
full-scale supernova simulations with the equation of state of
\citet{Lattimer91}, using the energy-dependent neutrino transport of the
\textsc{Vertex} code (cf. \citealt{Buras06b} and also Fig.~1 in
\citealt{Scheck06}).

In the simulations presented here we also explore the consequences of a
different time-dependence of the neutrino luminosities imposed at the inner
grid boundary (see Sects.~\ref{sec:1Dmodels} and \ref{sec:boundaryvariations}).
The explosion energy of a model is mostly determined by the choice of the
initial values of these luminosities (in particular those of $\nu_{\mathrm{e}}$ and
$\bar\nu_{\mathrm{e}}$).  These initial values are constrained by the prescribed total
loss of neutrino energy from the core during the proto-neutron star cooling,
$\Delta E_{\nu,\mathrm{core}}^{\mathrm{tot}}$, and by the total loss of lepton
number, $\Delta Y_{\mathrm{e,core}}$ (see Eqs.~D.13--D.16 in
\citealt{Scheck06}, where $L_{\nu}^{\mathrm{tot},0}t_{\mathrm{L}}$ in the last
equation should be replaced by $\Delta E_{\nu,\mathrm{core}}^{\mathrm{tot}}$).
The relative contribution of $\nu_{\mathrm{e}}$ to the total core luminosity is set to
20\%, (i.e., $K_{\nu_{\mathrm{e}}} = 0.2$ in terms of the parameters introduced in
\citealt{Scheck06}), the contribution of $\bar\nu_{\mathrm{e}}$ is determined from
requesting $\Delta Y_{\mathrm{e,core}} = 0.3$, and the muon and tau neutrino
contributions then follow from Eq.~(D.12) in \citet{Scheck06}. The mean
energies of the neutrinos entering the computational grid at the inner boundary
are chosen to be $\langle \epsilon_{\nu_{\mathrm{e}}}\rangle^{\mathrm{ib}} = 12\,$MeV,
$\langle \epsilon_{\bar\nu_{\mathrm{e}}}\rangle^{\mathrm{ib}} = 16\,$MeV, and $\langle
\epsilon_{\bar\nu_x}\rangle^{\mathrm{ib}} = 20\,$MeV when $\nu_x$ denotes muon
and tau neutrinos and antineutrinos. These energies are kept constant with
time.

Because of the contraction and postbounce accretion of the 
proto-neutron star, the density and optical depth in the layers
near the inner grid boundary can increase to such large values
that the application of our transport approximation becomes
inefficient by the required very fine zoning, and the equation
of state fails to describe the dense stellar matter. Whenever the
$\nu_{\mathrm{e}}$ optical depth begins to exceed a certain value (usually
chosen to be 300), we shift the inner boundary to a larger radius
$\widetilde{R}_{\mathrm{ib}}(t_{\mathrm{cut}})$ and thus to a larger
mass shell where the neutrino optical depth is significantly
lower (usually 200 for $\nu_{\mathrm{e}}$). The
additional excised baryonic mass is added to the previous
core mass and the gravitational mass of the new, increased core
is set equal to the gravitational mass computed at radius 
$\widetilde{R}_{\mathrm{ib}}(t_{\mathrm{cut}})$ where the new inner grid 
boundary is placed 
(see Sect.~\ref{sec:gravity}). The subsequent motion of the
new boundary for $t > t_{\mathrm{cut}}$ is assumed to follow the 
function
\begin{equation}
  R'_{\mathrm{ib}}(t) = R_{\mathrm{f}} +
  ( \widetilde{R}_{\mathrm{ib}}(t_{\mathrm{cut}}) - R_{\mathrm{f}} ) \,
  \exp \left[ v (t-t_{\mathrm{cut}}) / ( \widetilde{R}_{\mathrm{ib}}
  (t_{\mathrm{cut}}) - R_{\mathrm{f}}) \right] \,,
\label{eq:ibmotionew}
\end{equation}
where $v < 0$ is the recession velocity of the mass shell of the
new boundary at time $t_{\mathrm{cut}}$. The
new boundary contracts in a very similar way as the previous one
because the removed shell is very narrow. The neutrino luminosities
and mean energies of the streaming neutrinos imposed at the new 
boundary at $t = t_{\mathrm{cut}}$
are chosen to be the values computed with the transport scheme at
this radius and to have the same time behavior as the initial
boundary luminosities and mean energies.

\subsection{Gravity}
\label{sec:gravity}

Relativistic effects are taken into account in our Newtonian hydrodynamics code
by using an ``effective relativistic gravitational potential'' \citep{Rampp02}.
The simulations presented in this paper (different from those of
\citealt{Scheck06}) employ the improved version of this potential described by
\citet{Marek06}, who found excellent agreement with fully relativistic
calculations during core collapse and the first several hundred milliseconds
after core bounce (tests for the later neutrino-wind phase can be found in
Sect.~\ref{sec:relativity}).

According to \citet{Rampp02} and \citet{Marek06}, the 
relativistic equation of motion can be rearranged in a form similar
to the Newtonian one by replacing the Newtonian gravitational 
potential with a modified TOV potential:
\begin{equation}
\Phi_{\mathrm{TOV}}(r) = -4\pi G \int\limits_{r}^{\infty} 
        \frac{\mathrm{d}r'}{r'^2}
        \left(\frac{{\widetilde m}_{\mathrm{TOV}}}{4 \pi} + 
         {r'^{3}P\over c^2} \right)
        \frac{1}{\Gamma^{2}} \left( \frac{\rho c^2 +e+P}{\rho c^2} \right)
        \, ,
\label{eq:gravpot}
\end{equation}
where $\rho$ is the rest-mass density, $e=\rho \epsilon$ the internal energy
density with $\epsilon$ being the specific internal energy, and $P$ the gas
pressure. The usually rather small corrections of the gravitational potential
due to neutrino pressure, energy density, and flux terms (see 
\citealt{Marek06}) are neglected in Eq.~(\ref{eq:gravpot}).  
The ``modified TOV mass'' ${\widetilde m}_{\mathrm{TOV}}$ is given by
\begin{equation}
  {\widetilde m}_{\mathrm{TOV}}(r) = 4 \pi \int_{0}^{r} \mathrm{d}r' 
   r'^{2} \left(\rho + {e\over c^2}\right) \Gamma
\label{eq:tovmass}
\end{equation}
with the metric function 
\begin{equation}
  \Gamma = \sqrt{1 + {v^{2}\over c^2} -\frac{2 G 
                     {\widetilde m}_{\mathrm{TOV}}}{rc^2}}
  \ .
\label{eq:metricfunction}
\end{equation}
The extra factor $\Gamma$ in Eq.~(\ref{eq:tovmass}) compared to the
relativistic definition of the TOV mass enters the mass integral for reasons of
consistency with the Newtonian hydrodynamics equations and accounts for the
fact that in the Newtonian code there is no distinction between local proper
volume and coordinate volume, i.e., ${\mathrm{d}}{\cal V} = {\mathrm{d}}V$ (for
more details, see \citealt{Marek06}).

There is, however, an important difference of our calculations compared to
those performed by \citet{Marek06}. While the latter included the whole neutron
star down to the center, the use of the inner grid boundary at a radius
$R_{\mathrm{ib}} > 0$ in the present work prevents the evaluation of the
integral in Eq.~(\ref{eq:tovmass}) within the neutron star core. We solve this
problem by starting our calculations with a given value of the modified TOV
mass of the core at $t = 0$, ${\widetilde
  m}_{\mathrm{TOV}}(R_{\mathrm{ib}},0)$, which was provided to us as part of
the data set for the initial conditions of our simulations. For $t > 0$ we then
approximately evolve the modified TOV mass according to the expression
\begin{eqnarray}
{\widetilde m}_{\mathrm{TOV}}(R_{\mathrm{ib}},t) = 
     {\widetilde m}_{\mathrm{TOV}}(R_{\mathrm{ib}},0)
     &-&\! \int_{0}^{t} L_{\nu}^{\mathrm{ib}}(t')\,\mathrm{d}t' \nonumber \\
     &-&\! \int_{0}^{t}\! 4\pi R_{\mathrm{ib}}^2(t')P_{\mathrm{ib}}(t')
     {\mathrm{d}R_{\mathrm{ib}}\over {\mathrm{d}}t'} \, \mathrm{d}t' \, ,
\label{eq:tovmassib}
\end{eqnarray}
where the second term on the rhs yields the energy loss from
the neutron star core by the total neutrino luminosity at the
inner boundary, $L_{\nu}^{\mathrm{ib}}(t)$, and the last term 
represents the compression ($P\mathrm{d}V$) work done on the 
core at the contracting
inner boundary. The total modified TOV-mass at radius $r$, which
we consider as ``gravitational mass'', is thus given by
\begin{equation}
\widetilde{m}_{\mathrm{TOV}}(r,t) = \widetilde{m}_{\mathrm{TOV}}(R_{\mathrm{ib}},t) +
4\pi \int_{R_{\mathrm{ib}}}^{r} \mathrm{d}r' r'^{2} 
\left(\rho + {e\over c^2}\right)\Gamma
\ .
\label{eq:tovmassr}
\end{equation}

\begin{figure}[tpb!]
  \centering
  \includegraphics[width=8.5cm]{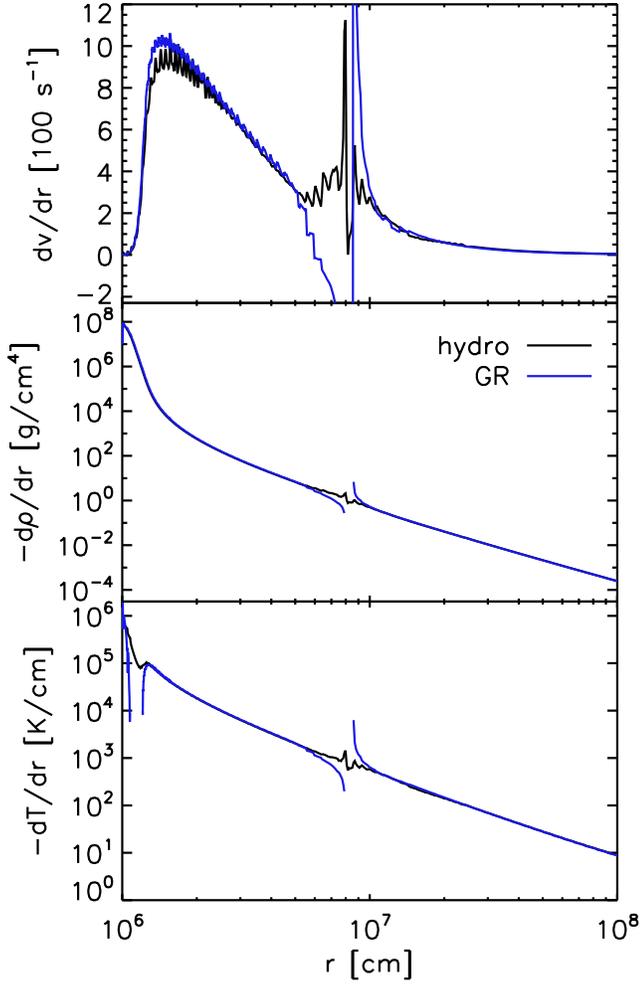}
  \caption{Derivatives of the velocity (top), density (middle), and
  temperature (bottom) as functions of radius from our hydrodynamic
  model M15-l1-r1 at 1.5$\,$s after bounce (black lines) compared to
  these derivatives as computed from the relativistic stationary 
  wind equations of Eqs.~(5)--(7) in \citet{Thompson01} 
  (blue curves). 
  These equations were evaluated by using the values of all 
  gas quantities as provided by our hydrodynamic model. Consistency
  between our hydrodynamics results (with approximative treatment of
  relativity) and the fully relativistic wind solution would require
  the corresponding lines to lie on top of each other. The agreement
  is very good and the sonic point is located at about 80$\,$km
  in both cases. This location is a critical point of the wind equations,
  which explains the pathological behavior of the curves there. The
  evaluation of the expression for the temperature gradient
  is numerically inaccurate in a region where the two terms
  in Eq.~(7) of Thompson et al.'s paper are very large and
  have opposite signs, in which case ${\mathrm{d}}T/{\mathrm{d}}r$
  becomes slightly positive while the hydrodynamical result is still
  negative.
  }
  \label{fig:dvdr}
\end{figure}

\begin{figure}[tpb!]
  \centering
   \includegraphics[width=7.35cm]{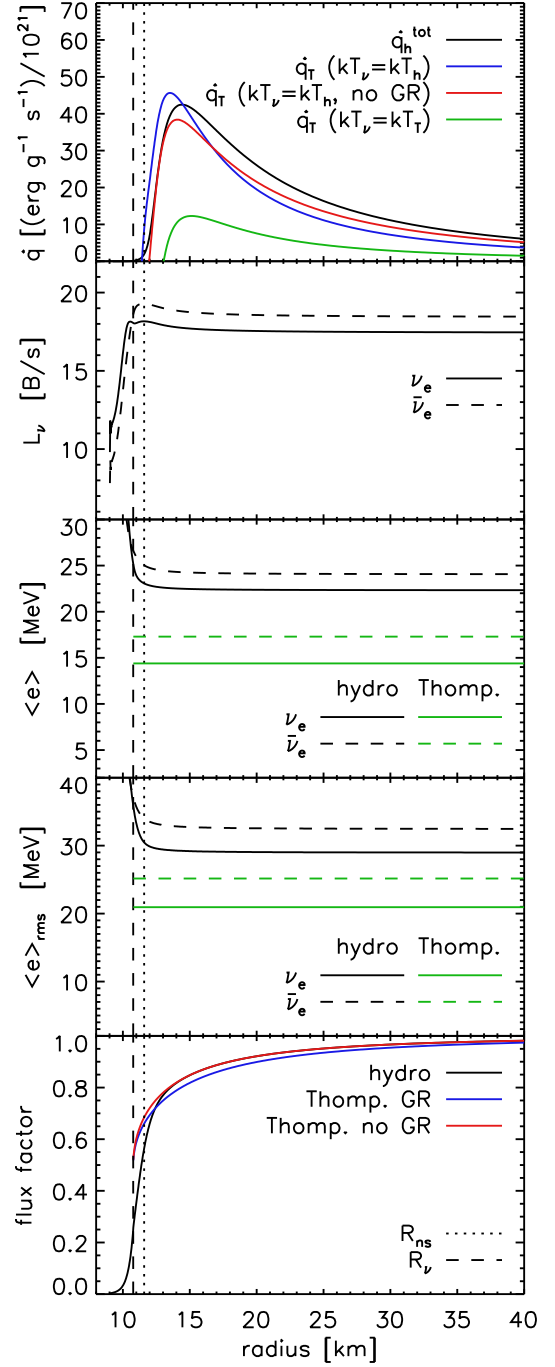}
  \caption{Radial profiles of the net total neutrino heating 
    rate $\dot q^{\mathrm{tot}}$ (top), $\nu_{\mathrm{e}}$ and $\bar\nu_{\mathrm{e}}$
    luminosities $L_\nu$ (second panel, in bethe per
    second or $10^{51}\,$erg$\,$s$^{-1}$), mean neutrino energies
    $\left\langle \epsilon_\nu\right\rangle$
    (third panel; Eq.~\ref{eq:ave}), rms energies 
    $\left\langle \epsilon_\nu\right\rangle_{\mathrm{rms}}$
    (fourth panel;
    Eq.~\ref{eq:rmsave}), and flux factor for model M15-l1-r1 at
    1.5$\,$s after bounce. For comparison with 
    Thompson at al.\ (2001), we also show 
    $\left\langle \epsilon_\nu\right\rangle$ and
    $\left\langle \epsilon_\nu\right\rangle_{\mathrm{rms}}$
    as used in that paper, rescaled to our 
    values of $L_{\nu_{\mathrm{e}}}$ and $L_{\bar\nu_{\mathrm{e}}}$, the flux factor
    with and without relativistic corrections from that work,
    and the Newtonian and GR charged-current heating plus
    cooling rates using Thompson et al.'s formulas,
    evaluated with our neutrino luminosities and rms energies
    (red and blue curves) or with the rescaled rms energies of
    Thompson et al.\ (green curve). The vertical dashed line
    marks the position of the $\nu_{\mathrm{e}}$-sphere, and the vertical 
    dotted line the neutron star ``surface'' at a density of
    $10^{11}\,$g$\,$cm$^{-3}$.
    }
  \label{fig:grneutrinos}
\end{figure}

\section{Comparison with fully relativistic wind solutions}
\label{sec:relativity}

In this section we will discuss our simulation approach in view of other
published work on relativistic steady-state solutions for neutrino-driven
winds. General relativistic (GR) effects have been recognized to cause
important changes of the neutrino wind properties, e.g., to lead to a decrease
of the expansion timescale and to an increase of the wind entropy, see
\citet{Qian96}, \citet{Cardall97}, \citet{Sumiyoshi00}. A comprehensive
discussion of these effects in comparison with the Newtonian treatment was
provided by \citet{Otsuki00} and \citet{Thompson01}.

In our simulations we account for relativistic gravity only by 
using a modified effective potential (Sect.~\ref{sec:gravity}), but
otherwise
we solve the Newtonian equations of hydrodynamics. We also ignore 
relativistic redshift and ray bending effects in our description 
of the neutrino transport \citep[cf.][]{Scheck06}.

The use of the generalized potential in a Newtonian
hydrodynamics code was shown previously to yield results in very 
good agreement with relativistic core-collapse simulations up 
to several 100$\,$ms after core bounce
\citep{Liebendorfer05,Marek06}. 
For the later neutrino-wind phase we tried to compare with
solutions plotted by \citet{Thompson01}
for cases when our neutron star masses, neutron star radii, 
and neutrino-heating
rates were similar to the ones considered in that paper. 
Unfortunately, we were unable to find moments in our simulations
where all relevant parameters match up exactly
the cases considered by \citet{Thompson01}. As far as a 
comparison was possible, we observed satisfactory agreement in the
main properties characterizing the wind. 

A more quantitative comparison is hampered by the fact that 
relativistic neutrino-wind simulations are not available to us. We
therefore decided to make use of Eqs.~(5)--(7) for the velocity 
derivative, $\partial v/\partial r$, the density derivative,
$\partial \rho/\partial r$, and the temperature derivative,
$\partial T/\partial r$, in the paper of
\citet{Thompson01}. Figure~\ref{fig:dvdr} shows these
derivatives as functions of radius at a certain time for
one of our models, compared to the results from Thompson et al.'s 
fully relativistic expressions. 
Evaluating the latter, we took all quantities on the rhs.\ of 
the formulas (velocity $v$, adiabatic sound speed 
$c_{\mathrm{s}}$, density $\rho$,
gravitational mass $M$, neutrino heating rate $\dot q$, etc.) from
our model. Ideally, the pairs of
corresponding curves in Fig.~\ref{fig:dvdr} should fall on top of
each other, which would demonstrate consistency
of both calculations. The overall agreement of the two cases is
very good, with a small difference being visible only around the 
maximum of the acceleration, which, however, is located at the 
same radius. Also the sonic point is
nearly at the same position of about 80$\,$km (we are not disturbed 
by the pathological behavior of the curves in this region,
where the expressions for the derivatives have a critical
point). We therefore conclude that our approach reproduces the
most important features of the relativistic solution, and that
relativistic kinematics (which we ignore) is of minor importance
compared to the effects of the stronger GR potential, which
makes the proto-neutron star more compact and the density and
temperature gradients in the neutrinospheric region steeper than
in Newtonian gravity.

We also compared our neutrino heating and cooling rates with
those used by \citet{Otsuki00} and \citet{Thompson01}.
Figure~\ref{fig:grneutrinos} shows the
radius-dependent net (i.e., heating minus cooling) specific 
rate of neutrino energy deposition by the $\beta$-processes
according to Eqs.~(20) and 
(21) of \citet{Thompson01} with and without corrections 
for relativistic redshift and ray bending, evaluated
at all radii with the stellar parameters and the neutrinospheric 
values of the $\nu_{\mathrm{e}}$ 
and $\bar\nu_{\mathrm{e}}$ luminosities and mean energies from one of our
simulations. The data were taken from the same model and time
used in Fig.~\ref{fig:dvdr}. The behavior of
both curves agrees qualitatively with Fig.~5a of 
\citet{Otsuki00}. Close to the neutrinosphere ray bending
effects enhance the net heating (since GR causes a reduction 
of the flux factor as visible for Thompson et al.'s prescription
of this quantity in Fig.~\ref{fig:grneutrinos}), whereas 
gravitational redshifting of the neutrino luminosities and
energies grows monotonically with distance from the neutrinosphere
and finally wins, reducing the GR rate below the Newtonian value.
\citet{Otsuki00} performed test calculations to disentangle
the influence of GR corrections in the neutrino
treatment from that of the relativistic terms in the wind 
structure equations. In spite of the sizable change of the local
heating rate, \citet{Otsuki00} found that neutrino redshift
and ray bending have only little impact on the wind entropy.
Similar conclusions were arrived at by \citet{Thompson01}.

In Fig.~\ref{fig:grneutrinos} also the total specific rate 
of neutrino heating and cooling from our hydrodynamical model
is displayed. This rate includes all contributing processes,
i.e. besides the $\beta$-reactions of $\nu_{\mathrm{e}}$ and $\bar\nu_{\mathrm{e}}$
absorption and production also energy transfer by 
the scattering off electrons, positrons, and free
nucleons, and neutrino-antineutrino pair annihilation, to which
neutrinos of all flavors contribute (cf.\ the
appendix of \citealt{Scheck06}). This total rate is similar to
the neutrino capture and emission rates of Eqs.~(20) and (21)
of \citet{Thompson01}, because for the considered situation
the neutrino luminosities are high (see Fig.~\ref{fig:grneutrinos})
and therefore the wind mass loss rate is large and the wind 
entropy fairly low ($s_{\mathrm{w}} \la 50\,k_{\mathrm{B}}$
per nucleon). At such conditions of high wind density and
modest abundance of $\mathrm{e}^{+}\mathrm{e}^{-}$-pairs, the other
reactions do not contribute significantly to the total
rate of energy deposition. 

We point out here that our approximative treatment of neutrino transport
evolves the transport solution self-consistently with the temperature and
density structure of the stellar medium. This is different from the light-bulb
approach of previous steady-state or hydrodynamical wind studies
\citep[e.g.,][]{Sumiyoshi00,Otsuki00,Thompson01}. Inside the neutrinosphere
neutrinos and matter are in equilibrium, around the neutrinosphere neutrinos
begin to decouple thermodynamically from the medium, and at some larger
distance they start streaming freely. A changing radial structure of the
contracting neutron star leads to changes of the neutrino luminosities and mean
energies, and the gradual loss of neutrinos drives the cooling and
deleptonization of the surface-near layers of the neutron star. In previous
wind studies (except full supernova models), such a coupling and
interdependence was ignored. Close to the neutron star surface the flux factor
(or flux dilution factor) used in our transport, which is based on a Monte
Carlo calibration by \citet{Janka91b}, is lower than the vacuum approximation
chosen by \citet{Thompson01} and \citet{Otsuki00}, see
Fig.~\ref{fig:grneutrinos}. \citet{Thompson01} have tested the improved
description by \citet{Janka91b} and found that its effects are negligible for the
range of model conditions considered by them. This, however, is true only
during phases where the density gradient near the neutrinosphere is very steep
and in regions where the neutrino luminosities have already reached their
asymptotic values.

The most important difference of our simulations compared to 
other relativistic wind studies is the different treatment
of the spectra in the neutrino transport. In our ``grey'' but 
non-equilibrium description of neutrino number and energy 
transport, we determine a neutrino spectral 
temperature that is independent of the matter temperature
and can be different from it (for details, see the appendix
of \citealt{Scheck06}). This leads to higher mean energies
of $\nu_{\mathrm{e}}$ and $\bar\nu_{\mathrm{e}}$ radiated from the neutrinosphere 
than considered in the other works.
Figure~\ref{fig:grneutrinos} shows these mean energies
as functions of radius, defined once as the ratio of neutrino 
energy flux to neutrino number flux,
\begin{equation}
\left\langle \epsilon_\nu\right\rangle \equiv 
\frac{L_{e}}{L_{n}} \ , 
\label{eq:ave}
\end{equation}
and another time as rms energy,
\begin{equation}
\left\langle \epsilon_\nu\right\rangle_{\mathrm{rms}}
\equiv k_{\mathrm{B}}T_\nu \sqrt{{{\cal F}_5(\eta_\nu)\over
{\cal F}_3(\eta_\nu)}}\ ,
\label{eq:rmsave}
\end{equation}
which is the energy which enters the calculation of the 
neutrino absorption rates on nucleons (cf., for example,
\citealt{Scheck06}). In Eq.~(\ref{eq:rmsave}),
$T_\nu$ and $\eta_\nu$ are the spectral temperature and 
degeneracy, assuming that the neutrino spectra have
Fermi-Dirac shape, in which case
\begin{equation}
{\cal F}_n(y) = \int_0^\infty {\mathrm{d}}x\, {x^n \over 
1 + \exp{(x-y)} } \ .
\end{equation}
For comparison, Fig.~\ref{fig:grneutrinos} also presents the corresponding mean
energies and rms energies as used by \citet{Thompson01}, appropriately scaled
by $L_\nu^{1/4}$ to account for the larger neutrino luminosities considered
here, and taking $\eta_\nu = 0$ for neutrinos and antineutrinos. The net
heating rate computed with these rms energies is significantly lower than the
heating rate from our hydrodynamic model (Fig.~\ref{fig:grneutrinos}).

At first glance, our mean 
energies for $\nu_{\mathrm{e}}$ and $\bar\nu_{\mathrm{e}}$ might appear on the large
side. One must, however, take into account that the mean energies
in our simulations are significantly lower in the first 
$\sim\,$0.5 seconds when the neutron star is still rather
extended, and only increase as it heats up during contraction.
They reach a maximum between one and two seconds after bounce 
to decrease afterwards as the proto-neutron star cools
(see Sect.~\ref{sec:1Dresults}). So the conditions plotted in
Fig.~\ref{fig:grneutrinos} correspond to a time when the
neutrino luminosities are still rather high and the mean
energies in this phase at their maximum. Moreover, one should
remember that we ignore gravitational redshifting in our 
transport. The redshift from the neutrinosphere at radius
$R_\nu$ to infinity after the contraction of
the neutron star can become quite significant,
$\sqrt{1 - 2GM/(R_\nu c^2)} = \sqrt{1 - R_{\mathrm{s}}/R_\nu}
\approx 0.7$...0.8 for 
${1\over 2} \ga R_{\mathrm{s}}/R_\nu \ga {1\over 3}$,
which reduces the mean energies for a distant observer by
typically 20--30\%. In our Newtonian transport treatment 
we prefer to use the higher neutrinospheric energies for
evaluating the neutrino heating, because the neutrino-wind 
properties are mostly determined by the heating 
just outside of the neutrinosphere, where it is also strongest.

\section{Results}
\label{sec:1Dresults}

In this section we will present the results of our time-dependent
hydrodynamic simulations. An overview of the computed models
will be given in Sect.~\ref{sec:1Dmodels}.
We will begin with the description of a reference case in
Sect.~\ref{sec:reference}. This model was computed
for a certain choice of the time-dependent contraction of the 
inner grid boundary and of the neutrino luminosities imposed at
this boundary. We will then demonstrate that basic features of 
the wind termination shock 
can be understood by simple analytic considerations
(Sect.~\ref{sec:analytic}). Next, we will discuss the neutrino
wind evolution in different progenitor stars
(Sect.~\ref{sec:progenitors}), and finally will
investigate the influence of variations of the conditions at the
inner boundary (Sect.~\ref{sec:boundaryvariations}).

\begin{table}
\caption{Model parameters of our spherically symmetric simulations.
The different models are characterized by the chosen contraction 
of the inner grid boundary, which is expressed in terms of the
final radius $R_{\mathrm{f}}$ and the exponential contraction
timescale $t_{0}$ (cf.\ Eq.~\ref{eq:ibmotion}). Different choices
of these values are indiced by the extensions ``r1'', ``r2'', etc.\
of the model names. In addition, different initial luminosities
of $\nu_{\mathrm{e}}$ plus $\bar\nu_{\mathrm{e}}$ (measured in bethe [B] = $10^{51}\,$erg
per second) are imposed at the inner grid boundary
in case of our standard luminosity behavior (constant until 1$\,$s
and then a $t^{-3/2}$ decay). These variations are reflected by the
extensions ``l1'', ``l2'', etc.\ in the model names. Moreover,
the time-dependence of the boundary luminosity has been modified
to a luminosity decay that is more rapid than in the standard
description (Eqs.~\ref{eq:lumtime} and
\ref{eq:lumtime2}; models with ``lt'' in their names).
}
\begin{tabular}{ccccc}
\hline
\hline
Model & Contraction & $L_{\nu_{\mathrm{e}}}^{\mathrm{ib}}+
L_{\bar\nu_{\mathrm{e}}}^{\mathrm{ib}}$ & Progenitor Mass \\
      &  ($R_{\mathrm{f}}$, $\,t_{0}$)  & [B/s]
       & $[M_{\odot}]$   \\
\hline
M15-l1-r1  & 9$\,$km;    $\,0.1\,$s & $52.5$ & 15 \\
M15-l1-r2  & 9$\,$km;    $\,0.2\,$s & $52.5$ & 15 \\
M15-l1-r5  & 11$\,$km;   $\,0.1\,$s & $52.5$ & 15 \\
M15-l1-r6  & 14$\,$km;   $\,0.1\,$s & $52.5$ & 15 \\
M15-l2-r1  & 9$\,$km;    $\,0.1\,$s & $38.6$ & 15 \\
M15-l3-r3  & 10$\,$km;   $\,0.1\,$s & $35.8$ & 15 \\
M15-lt2-r3 & 10$\,$km;   $\,0.1\,$s & $55.2$ & 15 \\
M15-lt1-r4 & 10.5$\,$km; $\,0.1\,$s & $55.8$ & 15 \\
M10-l1-r1  & 9$\,$km;    $\,0.1\,$s & $52.5$ & 10 \\
M10-l5-r3  & 10$\,$km;   $\,0.1\,$s & $30.3$ & 10 \\
M20-l1-r1  & 9$\,$km;    $\,0.1\,$s & $52.5$ & 20 \\
M20-l3-r3  & 10$\,$km;   $\,0.1\,$s & $35.8$ & 20 \\
M20-l4-r3  & 10$\,$km;   $\,0.1\,$s & $33.1$ & 20 \\
M25-l5-r4  & 10.5$\,$km; $\,0.1\,$s & $30.3$ & 25 \\
\hline
\end{tabular}
\label{tab:tabmodels}
\end{table}

\subsection{The computed models}
\label{sec:1Dmodels}

A list of computed 1D models with their characterizing parameters
is given in Table~\ref{tab:tabmodels}. We have
performed simulations for progenitor stars with 10.2$\,M_\odot$
(data provided by A.\ Heger, personal communication), 
15$\,M_\odot$ (model s15s7b2; \citealt{Woosley95}), 
20$\,M_\odot$ (model s20.0, \citealt{Woosley02}), and 
25$\,M_\odot$ (model s25a28, \citealt{Heger01}).
Pre-collapse profiles of the last three models are plotted
in the appendix of \citet{Buras06b}.
The progenitor cores were computed through core collapse and 
bounce with the 
neutrino-hydrodynamics code \textsc{Vertex} and provided to us as
initial conditions for the present studies
a few milliseconds after shock formation (A.\ Marek, personal
communication). Extensions of the model names (``r1'', ``r2'',....)
indicate different prescriptions for 
the contraction of the inner grid boundary, whose motion was varied
by choosing different values
of the final radius $R_{\mathrm{f}}$ and of the exponential contraction
timescale $t_{0}$ (Eq.~\ref{eq:ibmotion}). Larger numbers in this
sequence correspond to less quickly contracting or less compact
neutron stars. Moreover, we varied the 
sum of the $\nu_{\mathrm{e}}$ and $\bar\nu_{\mathrm{e}}$ luminosities imposed at the grid
boundary with respect to the initial value as well as time-dependence.
In most of the calculations the luminosities were chosen to be 
constant during the first
second of postbounce evolution and to decay proportional to 
$t^{-3/2}$ afterwards as in \citet{Scheck06}. Such models are
labelled by the extensions ``l1'', ``l2'', etc., with a higher
number meaning a lower initial value of the boundary luminosity.
In another set of
calculations the boundary luminosities were assumed to have a 
smoother time-dependence (with no jumps in the time derivative) and
in particular with less neutrino energy radiated at late postbounce
times. The luminosities were prescribed as
\begin{equation}
  L(t) = \cases { L_0\ ,  &if $\ t \leq 0.5\,$s\,; \cr
                  L_0 f(t)\ , & if $\ t > 0.5\,$s\,, \cr }
 \label{eq:lumtime}
\end{equation}
with 
\begin{equation}
f(t) =  { \exp \left[ -(t-0.5)^2 \right] + 
             b \left[\, 1+ (t-0.5)^n\, \right]^{-1} \over
             (1+b) }\, ,
\label{eq:lumtime2}
\end{equation}
where the time $t$ is measured in seconds.
The corresponding models can be recognized 
by the letters ``lt'' in their names. The parameter $n$ was set
to 1.5 in both cases, while $b = 0.2$ was used
for model M15-lt1-r4 and $b = 0.3$ for M15-lt2-r3.

A comparison of these models allows us to study the influence of
different contraction behavior of the nascent neutron star. The 
contraction determines the release of gravitational energy from the
mantle layers of the compact remnant. The accretion luminosity 
generated in the mantle adds to the core flux (given
by the imposed boundary condition) and has an influence on the
explosion timescale and explosion energy of a model and thus on the 
location of the mass cut and the baryonic mass of the neutron star. The
corresponding gravitational mass, which decreases when energy is lost
in neutrinos (Eq.~\ref{eq:tovmassib}), the radius of the neutron star,
and the luminosities and mean energies of the radiated neutrinos 
are crucial parameters that directly affect the neutrino-wind
properties as functions of time \citep[see][]{Qian96}.   

We note that the supernova models we 
study here do not permit us to change individually and
independently all parameters and conditions that affect the
neutrino wind properties and that determine the behavior of the 
wind termination shock. The wind depends, e.g., on the neutron 
star gravitational potential and thus on the neutron star mass.
The latter
becomes larger when the postbounce accretion phase lasts
longer and the explosion happens later, or when the progenitor
is more massive and therefore the iron core and postbounce
accretion rate are larger. More massive progenitors thus
tend to produce neutron stars with bigger masses. For this reason
one cannot disentangle the influence of the progenitor
structure on the wind termination shock from the effects of the
neutron star mass on the neutrino wind.

In order to structure the discussion, we decided to first 
describe basic features in case of a
$15\,M_\odot$ reference model, then to vary the
boundary conditions for this model, and finally to present 
the results for different progenitors.

\begin{figure*}[tpb!]
\sidecaption
\centering
\includegraphics[width=12cm]{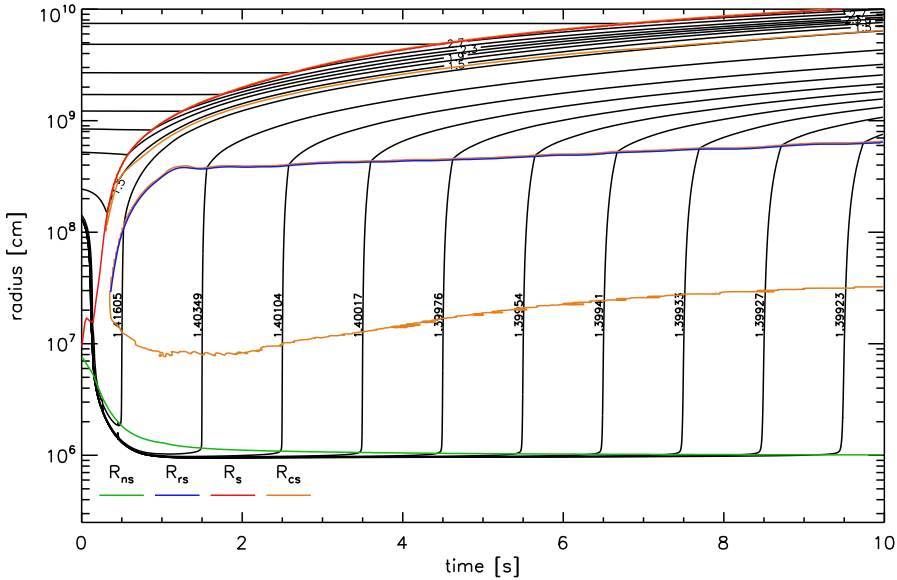} 
  \caption[]{Mass shell plot for the evolution of model M15-l1-r1.
  The explosion occurs about 0.2 seconds after bounce. The red line
  marks the supernova shock, the blue line the wind termination shock,
  the orange lines the locations where the expansion velocity of the
  gas equals the local sound speed (sonic points),
  and the green line the neutron star radius defined as the location
  where the density drops below $10^{11}\,$g$\,$cm$^{-3}$. A contact
  discontinuity separates the dense shell of ejecta that were 
  accelerated by the outgoing shock from the very dilute neutrino-driven
  wind. Mass shells in the wind are labelled by the corresponding 
  enclosed baryonic masses.}
  \label{fig:M15-l1-r1:massshells}
\end{figure*}

\begin{figure*}[!tpb]
   \centering
   \begin{tabular}{lr}
     \includegraphics[width=7cm]{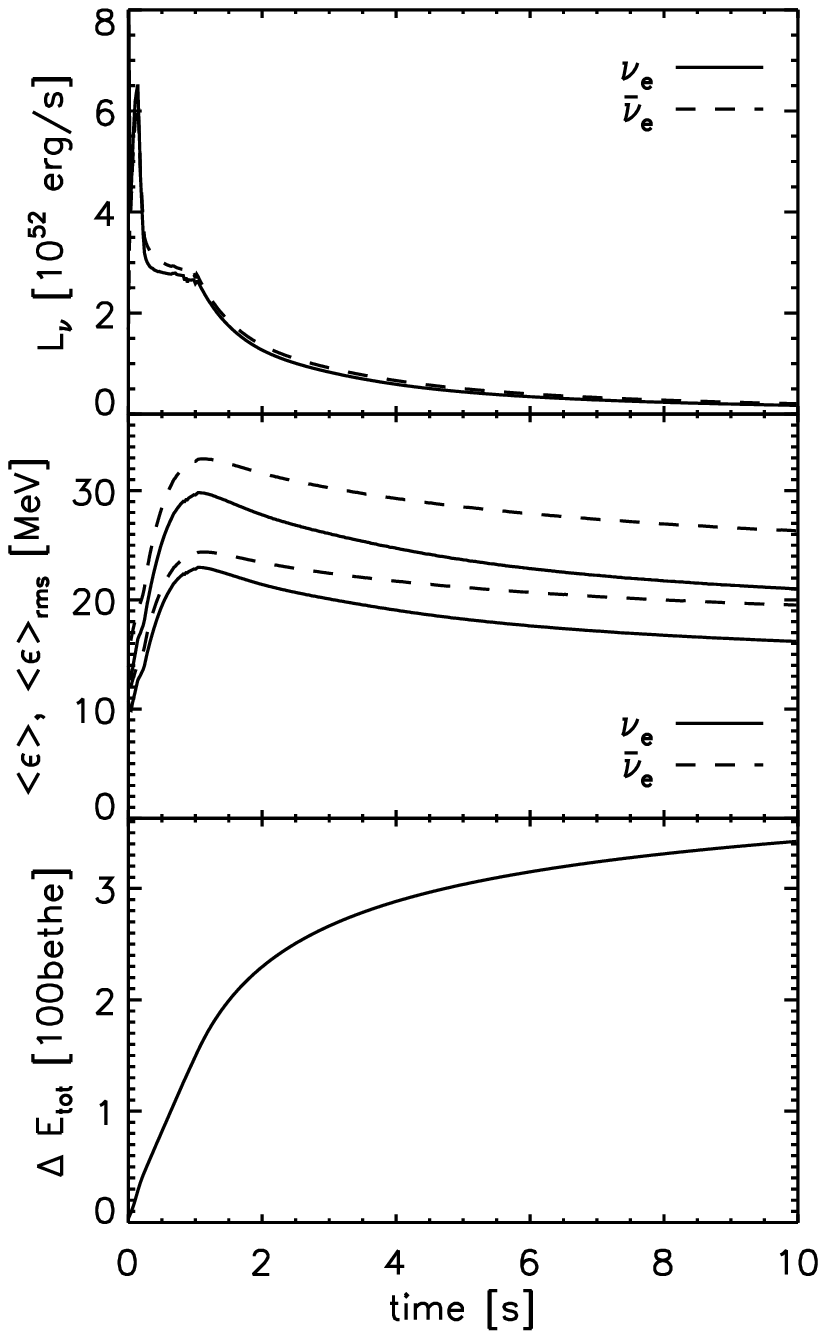} &
     \includegraphics[width=7cm]{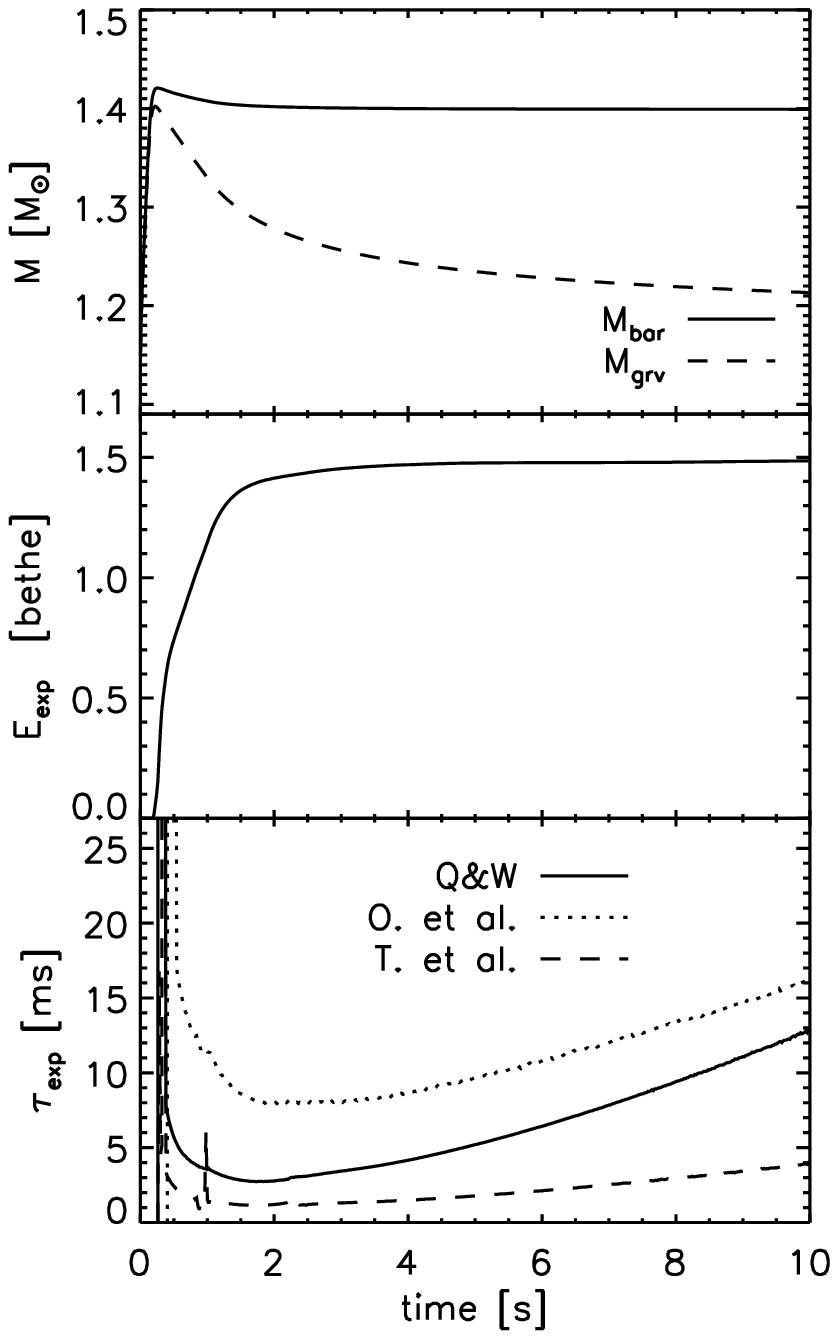} 
   \end{tabular}
   \caption{Luminosities, mean energies according to 
     Eq.~(\ref{eq:ave}), and rms energies (Eq.~\ref{eq:rmsave}, $\left\langle
       \epsilon\right\rangle_{\mathrm{rms}} > \left\langle
       \epsilon\right\rangle$) of $\nu_{\mathrm{e}}$ and $\bar\nu_{\mathrm{e}}$, and total energy
     radiated in neutrinos of all flavors for model M15-l1-r1 as functions of
     time (left), measured outside of the nascent neutron star (at a radius of
     500$\,$km). Note that we do not include gravitational redshifting in our
     neutrino treatment.  The rapid decline of the luminosities after about
     0.2$\,$ marks the end of the accretion phase of the forming neutron star
     at the onset of the explosion. The panels on the rhs side give the
     baryonic mass and the gravitational mass (Eq.~\ref{eq:tovmassr}) of the
     neutron star in model M15-l1-r1, the explosion energy, and the expansion
     timescales of the neutrino-driven wind as functions of time. For the
     latter, the results from three different definitions are displayed, namely
     those used by \citet{Qian96}, \citet{Otsuki00}, and \citet{Thompson01},
     given in Eqs.~(\ref{eq:tauqw}), (\ref{eq:tauotsuki}), and
     (\ref{eq:tauthompson}), respectively.}
   \label{fig:M15-l1-r1:neutrinos}
\end{figure*}

\begin{figure*}[!tpb]
   \centering
   \begin{tabular}{lr}
     \includegraphics[width=7cm]{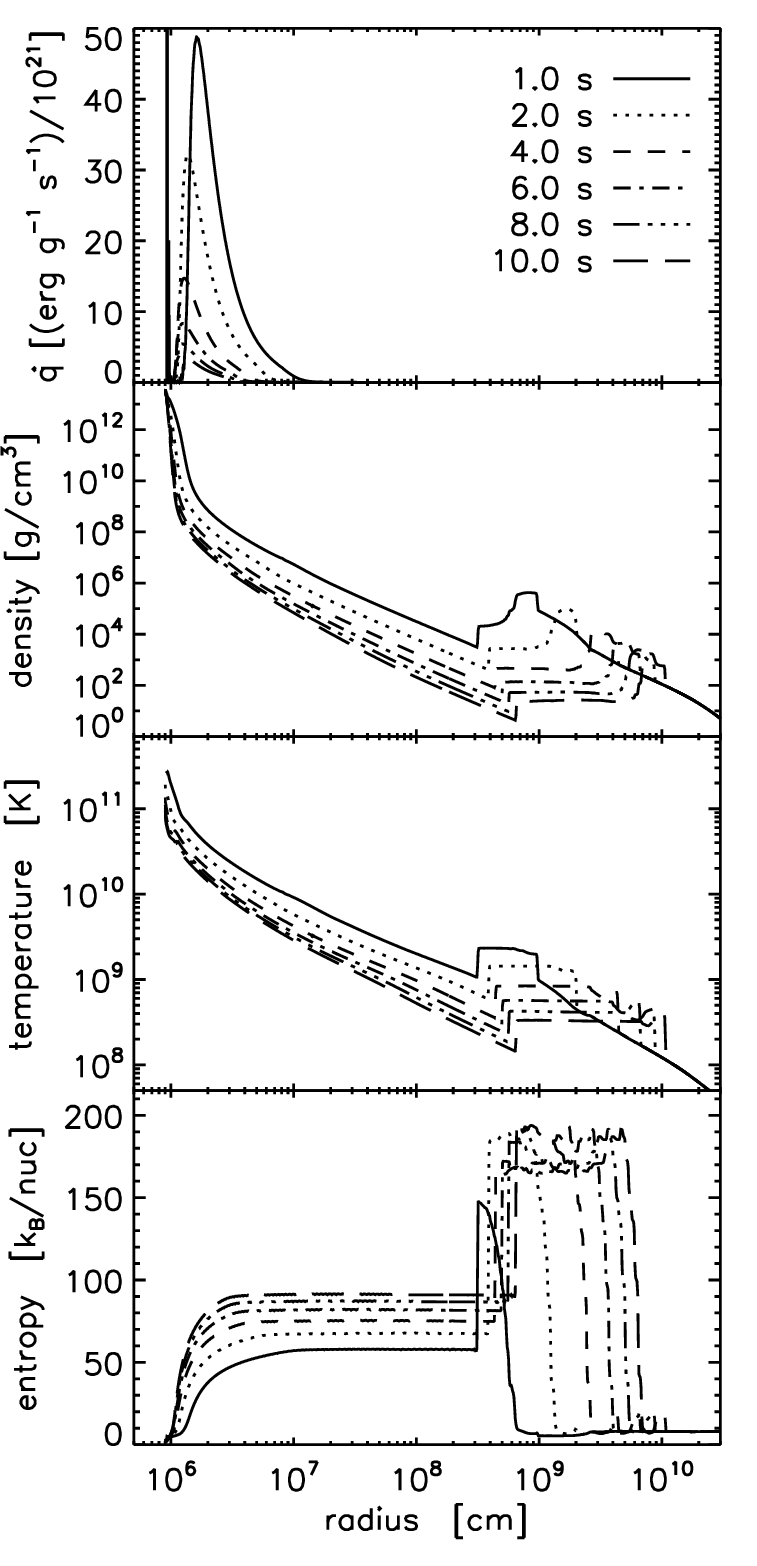} &
     \includegraphics[width=7cm]{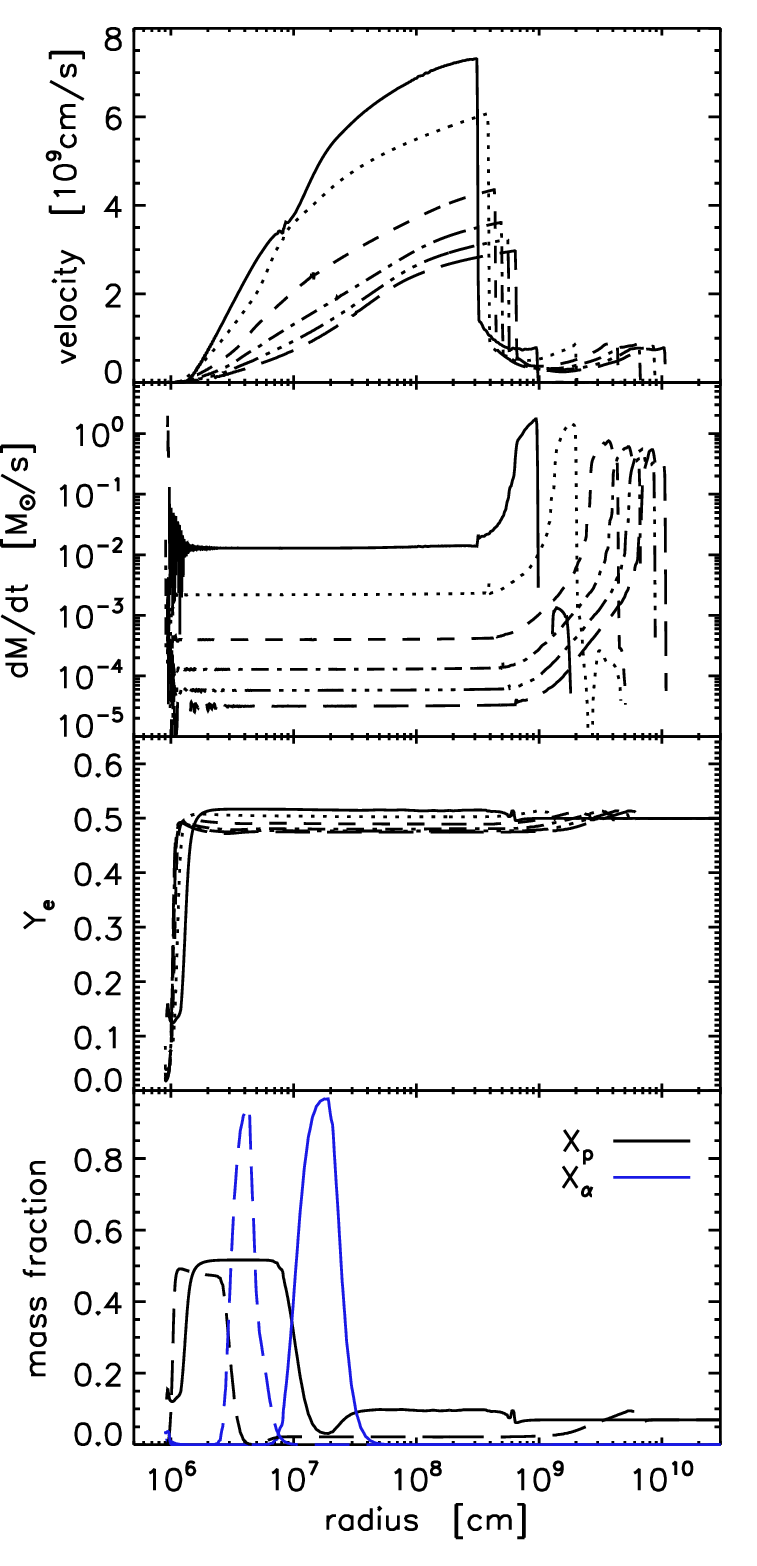}
   \end{tabular}
   \caption{Radial profiles of the net neutrino-heating rate $\dot q$,
    density, temperature, entropy (left, from top to 
    bottom), velocity, mass loss rate, electron fraction $Y_{\mathrm{e}}$, and
    mass fractions of free protons and $\alpha$ particles for the
    neutrino wind in model M15-l1-r1 at different postbounce times.
    For the mass fractions only the information for the first and last
    moments of time is provided. The wind termination shock is clearly
    visible in its effects on the velocity, density, temperature, and
    entropy of the outflow.}
   \label{fig:M15-l1-r1:profiles}
\end{figure*}

\begin{figure*}[!tpb]
   \centering
   \begin{tabular}{lr}
     \includegraphics[width=7cm]{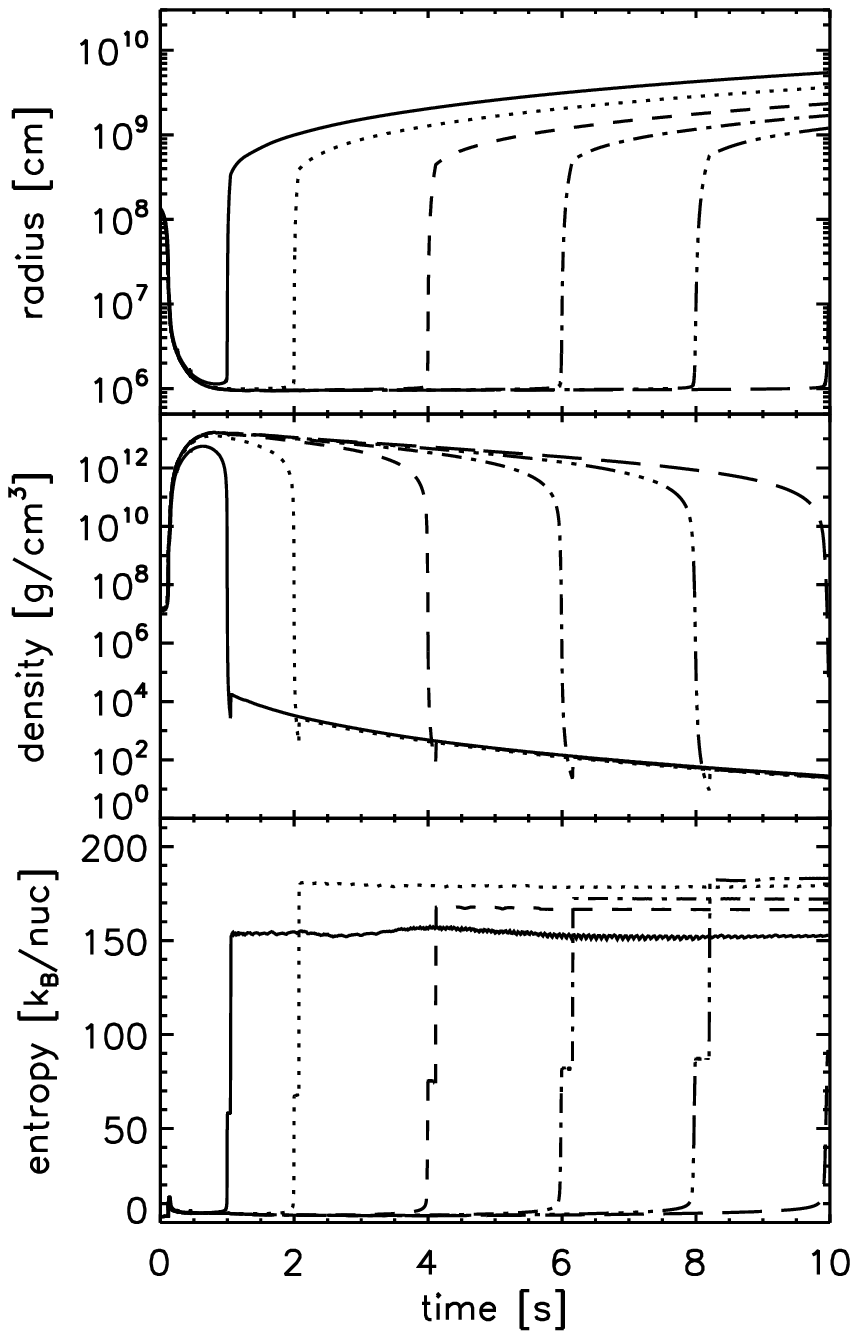} &
     \includegraphics[width=7cm]{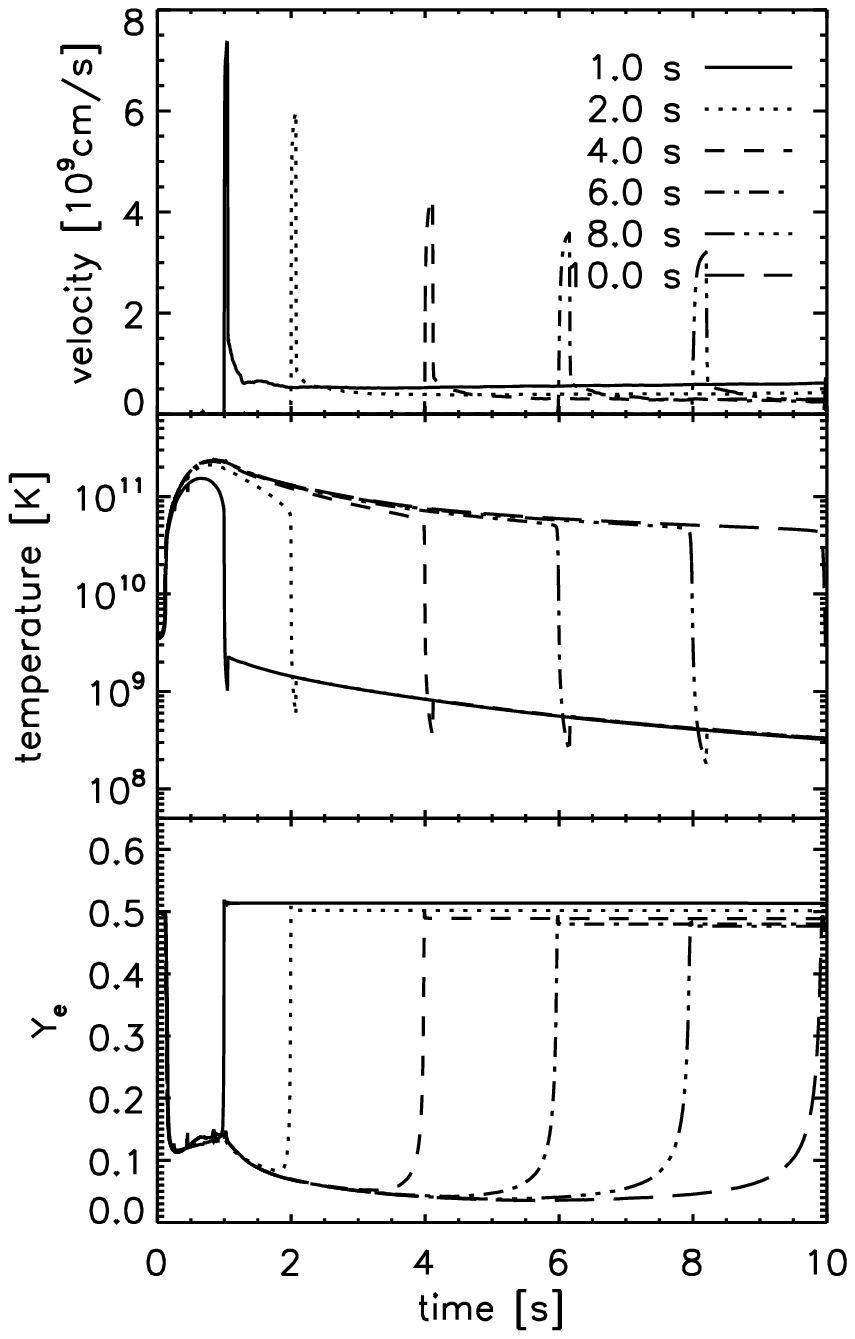}
   \end{tabular}
   \caption{Radius, density, entropy (left, from top to bottom),
    velocity, temperature, and electron fraction $Y_{\mathrm{e}}$ 
    as functions of time along the trajectories of different mass
    shells that are ejected in the neutrino-driven wind of 
    model M15-l1-r1. The times 
    correspond to the moments when the mass shells cross a radius of
    100$\,$km. After a very rapid expansion, the wind is abruptly
    decelerated by the termination shock. This leads to an increase
    of the entropy by more than a factor of two, and to a subsequently
    much slower decline of the temperature and density.}
   \label{fig:M15-l1-r1:trajectories}
\end{figure*}

\subsection{Evolution of a reference case}
\label{sec:reference}

In our reference model, M15-l1-r1, the conditions at the 
inner boundary were chosen such that the 15$\,M_\odot$ star
explodes with an energy of $\sim$1.5$\times 10^{51}\,$erg 
and the neutron star attains a final gravitational mass of
$\sim$1.2$\,M_\odot$ ($1.4\,M_\odot$ baryonic) and a radius
of 10$\,$km (Table~\ref{tab:finalresults}).
 
A mass-shell plot for the space-time evolution of this model is
given in Fig.~\ref{fig:M15-l1-r1:massshells}. The explosion sets 
in about 200$\,$ms after bounce (at the time $t_{\mathrm{exp}}$
given in Table~\ref{tab:finalresults}, which is defined as the 
moment when the total energy of expanding matter
starts to exceed $10^{49}\,$erg). At this time the stalled shock 
is revived by neutrino heating and starts continuous expansion 
with an average velocity of roughly 10.000$\,$km$\,$s$^{-1}$.
On its way out the shock reverses the infall of the swept-up
matter. After the onset of the explosion, ongoing neutrino energy 
transfer drives an outward acceleration of heated material in the
gain layer around the neutron star. At the interface between
this dilute neutrino-driven wind and the denser outer ejecta a
contact discontinuity is formed. 
Even farther behind the forward shock, the neutrino-driven
wind, whose velocity increases rapidly with distance from the
neutron star, collides with more slowly moving material and
is decelerated again. The strongly negative velocity gradient
at this location steepens into a reverse shock when the wind
velocity begins to exceed the local sound speed 
\citep{Janka95}. First indications of a forming wind
termination shock can be seen in Fig.~\ref{fig:M15-l1-r1:massshells}
at $t\ga\,$350$\,$ms post bounce at a radius $r\sim\,$300$\,$km.

Figure~\ref{fig:M15-l1-r1:neutrinos} displays the $\nu_{\mathrm{e}}$ and
$\bar\nu_{\mathrm{e}}$ luminosities and the mean energies emitted by
the nascent neutron star (gravitational redshift effects are
ignored). One can see the accretion phase with its production
of accretion luminosity ending at the time
the explosion sets in. The following plateau phase until
$t \approx 1\,$s and subsequent decay of the luminosities show the
influence of the time-dependence of the imposed boundary 
fluxes. This is also the case for the mean neutrino energies.
Their values increase during the first second of postbounce
evolution because the inner grid boundary and the
neutron star radius contract (Fig.~\ref{fig:M15-l1-r1:massshells}).
Consequently, the outer layers of the neutron star heat up due to
the conversion of gravitational energy to internal energy by
compression. After one second the rapid contraction is over
and the decay of the boundary luminosities leads to less energy
transport into these layers, which therefore begin to cool down,
causing the mean energies of the radiated neutrinos to decline.

Figure~\ref{fig:M15-l1-r1:neutrinos} also provides information
about the total energy carried away by neutrinos and antineutrinos
of all three lepton flavors, $\Delta E_{\mathrm{tot}}$, as a 
function of time, and the corresponding reduction of the 
gravitational mass of the nascent neutron star. The latter is
taken to be the modified TOV mass (Eq.~\ref{eq:tovmassr}) at the
neutron star radius $R_{\mathrm{ns}}$ (which is defined as the radius 
where the density is $10^{11}\,$g$\,$cm$^{-3}$). In contrast, the
baryonic mass of the neutron star, given by the rest mass enclosed
by the radius $R_{\mathrm{ns}}$, initially increases in the course
of accretion. After the explosion has been lauched,
it decreases again only slightly 
due to the mass loss in the neutrino-driven wind. 

The middle panel on the right side of
Fig.~\ref{fig:M15-l1-r1:neutrinos} reveals that only about 50\%
of the explosion energy (defined as the integral of the total
internal, kinetic, and gravitational energies in all zones where
the sum of these energies is positive) are carried by the 
neutrino-heated shell
of matter expanding right behind the shock after the onset of 
the explosion. The rest is contributed by the
early neutrino wind, and after 2$\,$s the energy has reached 95\%
of its final value. The panel below gives the expansion timescales
of the ejected mass shells in the neutrino wind. The first 
definition follows \citet[][Eq.~60]{Qian96}, who introduced
the dynamical timescale as
\begin{equation}
 \tau_{\mathrm{dyn}} =  \left. \frac{r}{v}\,
         \right|_{\,k_{\mathrm{B}}T=0.5\,\mathrm{MeV}} \ . 
\label{eq:tauqw}
\end{equation}
We compare this with the cooling timescale
used by \citet[][Eq.~23]{Otsuki00},
\begin{equation}
 \tau_T = \int_{\,k_{\mathrm{B}}T=0.5\,\mathrm{MeV}}
            ^{k_{\mathrm{B}}T=0.5\,\mathrm{MeV}/\mathrm{e}} 
        \frac{\mathrm{d}r}{v}  \ ,
\label{eq:tauotsuki}
\end{equation}
variations of which were considered by \citet[][cooling time between 
$T = 7\times 10^9\,$K and $T = 3\times 10^9\,$K]{Witti94} and 
\citet[][cooling time between $k_{\mathrm{B}}T = 0.5\,$MeV and
$k_{\mathrm{B}}T = 0.2\,$MeV]{Wanajo01}.  
The third definition we consider is the one of
\citet[][Eq.~32]{Thompson01}, who employed the e-folding time of
  the density instead of that of the temperature,
\begin{equation}
 \tau_\rho = \frac{1}{v} 
  \left|\frac{1}{\rho} \frac{\partial \rho}{\partial r}\, 
  \right|^{-1}_{\,k_{\mathrm{B}}T=0.5\,\mathrm{MeV}} \ ,
\label{eq:tauthompson}
\end{equation}
where we set for our Newtonian simulations $y=1$ in 
Thompson et al.'s Eq.~(32).
As can be expected from the fact that the wind is radiation-dominated
and therefore $s \propto T^3/\rho \sim\,$const, the
timescale $\tau_\rho$ is always significantly shorter than the
cooling timescale $\tau_T$ (Fig.~\ref{fig:M15-l1-r1:neutrinos}).
Ideally, in such a situation one would expect
$\tau_T/\tau_\rho = 3$, which is more closely reached at later stages
when the wind entropy is higher (Fig.~\ref{fig:M15-l1-r1:profiles}).
Due to the different mathematical expressions in
Eqs.~(\ref{eq:tauotsuki}) and (\ref{eq:tauthompson}), the factor
3 is never exactly realized. The third timescale, 
Eq.~(\ref{eq:tauqw}), yields a result that is between
the other two values during most of the computed postbounce
evolution and comes closer to the timescale of
Eq.~(\ref{eq:tauotsuki}) in the late stages of the simulations.

In Fig.~\ref{fig:M15-l1-r1:profiles} the radial profiles of
different wind quantities are given for our reference model
M15-l1-r1 at a number of postbounce times. The neutrino heating
accelerates the wind to a peak velocity of about 25\% of
the speed of light for neutrino luminosities $L_{\nu_{\mathrm{e}}}\approx
L_{\bar\nu_{\mathrm{e}}} \approx 3\times 10^{52}\,$erg$\,$s$^{-1}$. The
maximum velocity decreases as does the heating rate when the 
luminosities 
decline with time. The density and temperature in the wind 
region follow roughly the usual $r^{-3}$ and $r^{-1}$ behavior,
respectively, in the region where the entropy is a constant.
The profiles are slightly steeper and thus closer to these 
power laws at later times
when the wind entropy is higher and the wind therefore more
dominated by radiation pressure. The radial profiles of $\rho$
and $T$ also steepen at larger distance from the neutron star, 
leading to a visible increase of the
wind acceleration at the point where free nucleons recombine
to $\alpha$-particles and the neutrino heating 
ceases (see the corresponding panels in 
Fig.~\ref{fig:M15-l1-r1:profiles}). At this radius the entropy
of the outflow reaches its final value. 
This asymptotic wind entropy increases from 
about 60$\,k_{\mathrm{B}}$ per nucleon at 1$\,$s to around
90$\,k_{\mathrm{B}}$ at 10$\,$s. 

For (approximately) the same values of the $\nu_{\mathrm{e}}$ and $\bar\nu_{\mathrm{e}}$
luminosities, model M15-l1-r1 tends to yield somewhat lower 
expansion timescales, slightly lower entropies, and a bit higher mass
loss rates than those found by \citet{Thompson01}, see Figs.~5, 
8, and Tables~1 and 2 there. This can be understood on the
one hand by the smaller gravitational mass of the neutron star
in our model compared to the canonical 1.4$\,M_\odot$ star considered
by \citet{Thompson01}, and on the other hand it is caused by
our larger heating rates due to the higher mean 
neutrino energies (cf.\ our discussion in Sect.~\ref{sec:relativity}).
These differences affect the characteristic wind parameters with
different sensitivity.
According to \citet{Qian96}, the entropy scales with the 
neutrino luminosity $L$, the mean neutrino energy $\epsilon$, the
neutron star radius $R$, and the neutron star mass $M$ like
\begin{equation}
s\propto L^{-1/6}\epsilon^{-1/3}R^{-2/3}M\ , 
\label{eq:QWs}
\end{equation}
the expansion timescale like
\begin{equation}
\tau \propto L^{-1}\epsilon^{-2}R M \ , 
\label{eq:QWtau}
\end{equation}
and the wind mass loss rate like
\begin{equation}
\dot M \propto L^{5/3}\epsilon^{10/3}R^{5/3}M^{-2}
\label{eq:QWdotM}
\end{equation}
(modifications of these relations due to relativistic effects were
addressed by \citealt{Thompson01}).

\begin{table*}
\caption{Results of the 1D models at the end of the simulations
at $t = 10\,$s after bounce. $M_{\mathrm{bar}}$ is the baryonic
mass of the neutron star, $M_{\mathrm{grv}}$ its gravitational
mass (Eq.~\ref{eq:tovmassr}). Both masses are computed for the
matter inside the neutron star radius $R_{\mathrm{ns}}$. This radius
is defined as the location where the density is $10^{11}\,$g$\,$cm$^{-3}$.
$M_{\mathrm{eff}}$ denotes an ``effective mass'' of the neutron star,
for which a Newtonian force equals to the gravitational force associated with 
the modified TOV potential of Eq.~(\ref{eq:gravpot}) at radius $R_{\mathrm{ns}}$.
$\Delta E_{\mathrm{tot}}$ is the total energy radiated in neutrinos
of all flavors (measured in bethe [B] = $10^{51}\,$erg), $L_{\nu_{\mathrm{e}}}$
and $L_{\bar\nu_{\mathrm{e}}}$ are the luminosities of electron neutrinos and
antineutrinos measured at 500$\,$km (without gravitational 
redshifting), $\langle\epsilon_{\nu_{\mathrm{e}}}\rangle$ and
$\langle\epsilon_{\bar\nu_{\mathrm{e}}}\rangle$ are the corresponding mean energies,
$E_{\mathrm{exp}}$ is the explosion energy (note that this energy can
still decrease somewhat after 10$\,$s because of the negative binding
energy of the outer stellar layers, which is not included in the
given numbers), $t_{\mathrm{exp}}$ is the
postbounce time when the explosion sets in (defined as the moment
when the energy of expanding postshock matter exceeds $10^{49}\,$erg),
$s_{\mathrm{w}}$ is the asymptotic wind entropy per nucleon, and
$s_{\mathrm{rs}}$ the entropy of the outflow after its deceleration in
the wind termination shock.
}
\setlength\tabcolsep{5pt}
\begin{tabular}{cccccccccccccccc}
\hline
\hline
Model & time & $M_{\mathrm{bar}}$ & $M_{\mathrm{grv}}$ & $M_{\mathrm{eff}}$ & $\Delta E_{\mathrm{tot}}$ & $R_{\mathrm{ns}}$ & $L_{\nu_{\mathrm{e}}}$ & $L_{\overline{\nu}_{\mathrm{e}}}$ & $\langle \epsilon_{\nu_{\mathrm{e}}}\rangle$ & $\langle \epsilon_{\overline{\nu}_{\mathrm{e}}}\rangle$ & $E_{\mathrm{exp}}$ & $t_{\mathrm{exp}}$ & $s_{\mathrm{wind}}$ & $s_{\mathrm{rs}}$ \\ 
 & $[\mathrm{s}]$ & $[M_{\odot}]$ & $[M_{\odot}]$ & $[M_{\odot}]$ & $[100 \mathrm{B}]$ & $[\mathrm{km}]$ & $[\mathrm{B}/\mathrm{s}]$ & $[\mathrm{B}/\mathrm{s}]$ & $[\mathrm{MeV}]$ & $[\mathrm{MeV}]$ & $[\mathrm{B}]$ & $[\mathrm{s}]$ & $[k_{\mathrm{B}}/\mathrm{nuc}]$ & $[k_{\mathrm{B}}/\mathrm{nuc}]$ \\ 
\hline
M15-l1-r1 &    10.0 &   1.399 &   1.207 &   1.910 &   3.422 &   10.09 &    1.73 &    2.06 &   16.19 &   19.52 &   1.486 &   0.201 &   91.10 &  190.55 \\ 
M15-l1-r2 &    10.0 &   1.446 &   1.263 &   2.031 &   3.335 &   10.27 &    1.96 &    2.28 &   16.37 &   19.50 &   1.174 &   0.341 &   94.19 &  191.29 \\ 
M15-l1-r5 &    10.0 &   1.394 &   1.208 &   1.703 &   3.307 &   12.94 &    1.78 &    2.11 &   14.57 &   17.62 &   1.371 &   0.221 &   71.97 &  131.94 \\ 
M15-l1-r6 &    10.0 &   1.440 &   1.258 &   1.648 &   3.274 &   16.71 &    1.61 &    1.93 &   13.04 &   15.98 &   1.043 &   0.241 &   61.04 &   83.63 \\ 
M15-l2-r1 &    10.0 &   1.473 &   1.280 &   2.116 &   3.451 &    9.94 &    1.74 &    2.12 &   16.46 &   20.30 &   1.019 &   0.381 &  100.14 &  193.14 \\ 
M15-l3-r3 &    10.0 &   1.545 &   1.341 &   2.121 &   3.709 &   11.22 &    1.99 &    2.34 &   15.88 &   18.97 &   0.709 &   0.701 &   92.75 &  155.84 \\ 
M15-lt2-r3 &    10.0 &   1.397 &   1.253 &   1.906 &   2.602 &   11.28 &    1.33 &    1.54 &   15.04 &   18.12 &   1.239 &   0.221 &   90.19 &  132.57 \\ 
M15-lt1-r4 &    10.0 &   1.395 &   1.260 &   1.878 &   2.421 &   11.81 &    0.96 &    1.12 &   14.33 &   17.40 &   1.231 &   0.221 &   91.02 &   96.92 \\ 
M10-l1-r1 &    10.0 &   1.314 &   1.132 &   1.721 &   3.251 &   10.22 &    1.63 &    1.96 &   15.85 &   19.06 &   1.247 &   0.321 &   83.48 &  476.93 \\ 
M10-l5-r3 &    10.0 &   1.344 &   1.187 &   1.745 &   2.817 &   11.49 &    1.47 &    1.72 &   14.98 &   18.04 &   0.716 &   0.421 &   80.94 &  353.32 \\ 
M20-l1-r1 &    10.0 &   1.422 &   1.233 &   1.973 &   3.388 &   10.10 &    1.74 &    2.05 &   16.23 &   19.52 &   1.486 &   0.181 &   94.18 &  127.89 \\ 
M20-l3-r3 &    10.0 &   1.595 &   1.411 &   2.310 &   3.383 &   11.10 &    1.60 &    1.90 &   15.66 &   18.79 &   0.375 &   0.761 &  105.10 &    --- \\ 
M20-l4-r3 &    10.0 &   1.523 &   1.332 &   2.118 &   3.437 &   11.03 &    1.61 &    2.00 &   15.69 &   20.00 &   0.847 &   0.421 &   95.32 &  106.84 \\ 
M25-l5-r4 &    10.0 &   1.971 &   1.657 &   2.944 &   5.924 &   11.56 &    2.95 &    3.58 &   16.88 &   20.01 &   1.700 &   0.401 &  113.75 &  117.89 \\ 
\hline
\end{tabular}


\label{tab:finalresults}
\end{table*}

During the first $\sim\,$2 seconds after the onset of the explosion,
the neutrino wind is p-rich, i.e.\ $Y_{\mathrm{e}} > 0.5$. This is in agreement
with explosion models that employ a Boltzmann solver for the spectral
neutrino transport \citep[see][]{Buras06a,Pruet05}. 
Afterwards the electron fraction drops below 0.5, and gradually 
the wind develops increasing neutron excess. Qualitatively, this
trend to lower $Y_{\mathrm{e}}$ at later times is reproduced when
the neutrino luminosities and mean energies from the simulation are
inserted into the simple analytic relation $Y_{\mathrm{e}} \sim [\,1 +
(L_{\bar\nu_{\mathrm{e}}}\epsilon_{\bar\nu_{\mathrm{e}}}/L_{\nu_{\mathrm{e}}}\epsilon_{\nu_{\mathrm{e}}})\,]^{-1}$,
although the values do not agree quantitatively. We emphasize here
that the gray and approximative treatment of the neutrino transport 
employed in this work (for a critical assessment,
see \citealt{Scheck06}) is also not able to yield reliable results
for the electron fraction in terms of absolute numbers. The competition
of $\nu_{\mathrm{e}}$ and $\bar\nu_{\mathrm{e}}$ absorption on free neutrons and protons
sensitively determines the asymptotic value of $Y_{\mathrm{e}}$, an accurate
calculation of which requires detailed information of the neutrino
and antineutrino spectra in the comoving frame of the expanding
wind matter. The wind at late times might therefore become 
significantly more neutron rich than predicted in our models.
Figure~\ref{fig:M15-l1-r1:profiles} also reveals that the mass
loss rate reaches its asymptotic value closest to the 
neutrinosphere, and only then $Y_{\mathrm{e}}$ and finally the entropy
reach their asymptotic values.

At a radius of a few 1000$\,$km, the supersonic wind is abruptly
decelerated down in the termination shock. The compression leads to a
density and temperature increase. The conversion of kinetic to 
internal energy in the shock boosts the entropy to more than
twice the wind entropy in model M15-l1-r1. This is a much more
extreme impact of the termination shock than previously 
suggested in the literature
\citep{Thompson01}. The decelerated wind
material is accumulated in a dense shell between the forward and
reverse shocks. The pressure across this dense shell is nearly
uniform, while the contact discontinuity between the accumulated
wind matter and the dense layer of shock-accelerated progenitor
gas is clearly visible in the density profiles. 
One should also notice that the conditions at the wind termination
shock are by no means time-independent as previously assumed in 
nucleosynthesis calculations \citep[e.g.,][]{Wanajo02}.
Temperature and density at the reverse shock in model M15-l1-r1
evolve, because the radial position of the reverse shock as well
as the wind properties change with time. The impact of the wind
termination shock on the conditions in the expanding wind mass
shells is better visible in Fig.~\ref{fig:M15-l1-r1:trajectories},
where the time-evolution of different quantities is depicted as
seen comoving with some selected mass shells.
The extremely rapid decline of the temperature and density in
the fast wind are stopped and switch over to a much slower 
evolution. After the wind material has been added to the
dense shell between the two shocks, it moves with nearly constant
velocity. Its density therefore decays
approximately like $\rho\propto t^{-2}$ and because the gas is
radiation-dominated, its temperature follows roughly the power law
$T\propto t^{-2/3}$.

\subsection{Analytic discussion of the wind termination shock}
\label{sec:analytic}

The behavior of the wind termination shock and its effects on
the neutrino-driven outflow can basically be understood by
simple analytic considerations. For this purpose we consider
the three Rankine-Hugoniot shock jump conditions for mass,
momentum and energy flow,
\begin{equation}
 \rho_{\mathrm{rs}} u_{\mathrm{rs}} =  
 \rho_{\mathrm{w}} u_{\mathrm{w}}\ , 
\label{eq:shjump1}
\end{equation}
\begin{equation}
 P_{\mathrm{rs}}  + \rho_{\mathrm{rs}} u_{\mathrm{rs}}^{2}  
 = P_{\mathrm{w}}  + \rho_{\mathrm{w}} u_{\mathrm{w}}^{2} \ ,
\label{eq:shjump2}
\end{equation}
\begin{equation}
 \frac{1}{2}\,u_{\mathrm{rs}}^{2} + \omega_{\mathrm{rs}}  
 = \frac{1}{2}\,u_{\mathrm{w}}^{2} + \omega_{\mathrm{w}} \ ,
\label{eq:shjump3}
\end{equation}
where the indices $\mathrm{w}$ and $\mathrm{rs}$ denote 
quantities of the wind just ahead of the shock and of the shocked
matter just behind the shock, respectively. The fluid velocities 
$u_{\mathrm{w}}$ and $u_{\mathrm{rs}}$ are measured relative to
the shock velocity, $u = v - U_{\mathrm{s}}$, $P$ is the pressure,
$\rho$ the mass density, and $\omega = (\varepsilon + P)/\rho$
the enthalpy per mass unit when $\varepsilon$ is the internal
energy density of the gas. In case of radiation-dominated 
and nondegenerate conditions, one can write
$s = (\varepsilon + P)/(n_{\mathrm{B}} k_{\mathrm{B}}T)$
for the dimensionless entropy normalized by the baryon density 
$n_{\mathrm{B}} \rho/m_{\mathrm{B}}$ ($m_{\mathrm{B}}$ is the average baryon mass),
and therefore one gets
\begin{equation}
  s_{\mathrm{rs}} k_{\mathrm{B}}T_{\mathrm{rs}}  - s_{\mathrm{w}} 
k_{\mathrm{B}}T_{\mathrm{w}}
  = \frac{1}{2}m_{\mathrm{B}}(u_{\mathrm{w}}^{2} - u_{\mathrm{rs}}^{2}) \ .
\label{eq:shjump3a}
\end{equation}
Since the wind termination shock strongly decelerates the wind, 
the postshock and preshock velocities fulfill the relation 
$u_{\mathrm{w}}^2 \gg u_{\mathrm{rs}}^2$.
Thus the postshock entropy is approximately given by
\begin{equation}
s_{\mathrm{rs}}\approx
s_{\mathrm{w}}\,{k_{\mathrm{B}}T_{\mathrm{w}}\over 
                 k_{\mathrm{B}}T_{\mathrm{rs}}     } + 
\frac{1}{2}
\frac{m_{\mathrm{B}}u_{\mathrm{w}}^{2}}{k_{\mathrm{B}}T_{\mathrm{rs}}}\ .
\label{eq:entropy1}
\end{equation}
Again making the assumption that the gas on both sides of the
shock is radiation 
dominated, the dimensionless entropy per nucleon is given by
\begin{equation}
s = f_\gamma a_\gamma {(k_{\mathrm{B}}T)^3\over n_{\mathrm{B}}} \ ,
\label{eq:entropy2}
\end{equation}
where
$a_\gamma = a/k_{\mathrm{B}}^4 = 2.08\times 10^{49}\,$erg$^{-3}$cm$^{-3}$ 
is related to the radiation constant $a$, and $f_\gamma$ is a 
factor whose exact value depends on the temperature and thus the
mixture of radiation and $\mathrm{e}^{+}\mathrm{e}^{-}$-pairs; assuming zero electron
degeneracy, the corresponding range of values is
${4\over 3}\le f_\gamma \le {11\over 3}$. Equation~(\ref{eq:entropy2})
can be used to express $k_{\mathrm{B}}T$ ahead of and behind the shock
in terms of $s$ and $\rho$. Using also that the densities are
connected by 
$\rho_{\mathrm{rs}} = \beta\rho_{\mathrm{w}}$ with $\beta\sim 7$ for
a strong shock and radiation-dominated conditions, one derives
\begin{equation}
s_{\mathrm{rs}}  \approx  \left\lbrack \,
{s_{\mathrm{w}}^{4/3}\over \beta^{1/3}}
\,+\,\alpha^{1/3} {u_{\mathrm{w}}^2\over 
\rho_{\mathrm{w}}^{1/3}}\,\right\rbrack^{3/4} 
 \approx 
\left\lbrack \, {s_{\mathrm{w}}^{4/3}\over \beta^{1/3}} \,+\,
33.5\, {u_{\mathrm{w},9}^2\over \rho_{\mathrm{w},2}^{1/3}}\,
\right\rbrack^{3/4} ,
\label{eq:entropy3}
\end{equation}
where $\alpha \equiv f_\gamma a_\gamma m_{\mathrm{B}}^4/(8\beta)$,
$u_{\mathrm{w},9}$ is the wind velocity measured in 
$10^9\,$cm$\,$s$^{-1}$, $\rho_{\mathrm{w},2}$ the wind density
in 100$\,$g$\,$cm$^{-3}$, and the numerical value in the second 
expression was calculated with $\beta = 7$ and $f_\gamma = {4\over 3}$.
Equation~(\ref{eq:entropy3}) can be rewritten in terms of the wind
mass loss rate $\dot M_{\mathrm{w}}$ and reverse shock radius 
$R_{\mathrm{rs}}$, using 
\begin{equation}
\dot M_{\mathrm{w}} = 4\pi R_{\mathrm{rs}}^2\,\rho_{\mathrm{w}} 
v_{\mathrm{w}} 
\label{eq:mdotwind}
\end{equation}
and assuming that the shock velocity is negligible,
and therefore $v_{\mathrm{w}} = u_{\mathrm{w}}$, which gives
\begin{eqnarray}
s_{\mathrm{rs}} & \approx & \left\lbrack \,
{s_{\mathrm{w}}^{4/3}\over \beta^{1/3}}
\,+\,(4\pi\alpha)^{1/3} {R_{\mathrm{rs}}^{2/3} 
u_{\mathrm{w}}^{7/3}\over
\dot M_{\mathrm{w}}^{1/3}}\,\right\rbrack^{3/4}  \nonumber \\
& \approx &
\left\lbrack \, {s_{\mathrm{w}}^{4/3}\over \beta^{1/3}} \,+\,
28.7\, {R_{\mathrm{rs},8}^{2/3}
u_{\mathrm{w},9}^{7/3}\over \dot M_{\mathrm{w},-5}^{1/3}}\,
\right\rbrack^{3/4} .
\label{eq:entropy4}
\end{eqnarray}
Here $R_{\mathrm{rs},8}$ is in units of $10^8\,$cm and 
$\dot M_{\mathrm{w},-5}$ is normalized to $10^{-5}\,M_\odot$.
If the wind entropy is low, $s_{\mathrm{w}} \ll s_{\mathrm{rs}}$,
only the second terms in Eqs.~(\ref{eq:entropy3}) and 
(\ref{eq:entropy4}) are relevant.

\begin{figure}[tpb!]
  \centering
   \includegraphics[width=7.35cm]{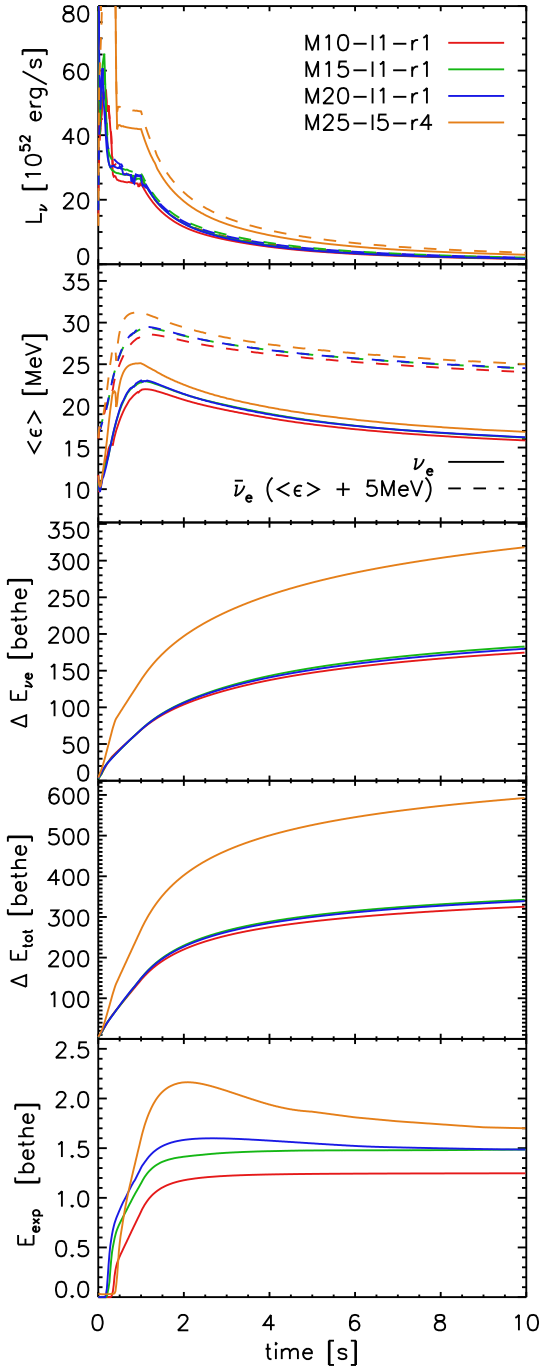}
  \caption{For a set of simulations with different progenitor
    stars, M10-l1-r1, M15-l1-r1, M20-l1-r1, and M25-l5-r4,
    the different panels show as functions of time
    (from top to bottom):
    The radiated luminosities of $\nu_{\mathrm{e}}$ and 
    $\bar\nu_{\mathrm{e}}$, the
    mean energies and rms energies of these neutrinos (all
    measured at a distance of 500$\,$km, disregarding
    gravitational redshift effects), the cumulative energy
    emitted in $\nu_{\mathrm{e}}$ and $\bar\nu_{\mathrm{e}}$, 
    $\Delta E_{\nu_{\mathrm{e}}}$,
    the total energy released in neutrinos and antineutrinos
    of all flavors, $\Delta E_{\mathrm{tot}}$, and the explosion
    energy of the models. } 
  \label{fig:prog-neutrinos}
\end{figure}

It is also possible to obtain an estimate of the reverse 
shock position from known supernova and wind parameters. 
In case of a strong shock, i.e., $P_{\mathrm{rs}} \gg
P_{\mathrm{w}}$, one can derive from Eqs.~(\ref{eq:shjump1}) 
and (\ref{eq:shjump2}) the relation 
$P_{\mathrm{rs}} \approx (1- \beta^{-1})
\rho_{\mathrm{w}}u_{\mathrm{w}}^2$. Using again
Eq.~(\ref{eq:mdotwind}) for $\rho_{\mathrm{w}}$, one gets
\begin{equation} 
R_{\mathrm{rs}} \approx \sqrt{\,\left( 1-{1\over \beta}\right)\,
{\dot M_{\mathrm{w}} u_{\mathrm{w}}\over 4\pi P_{\mathrm{rs}}}}\ .
\label{eq:rsradius1}
\end{equation}
Assuming the spherical shell between the forward shock at radius 
$R_{\mathrm{s}}\gg R_{\mathrm{rs}}$ and the reverse shock to
have constant pressure and to be radiation dominated, one
can make the approximation
\begin{equation}
P_{\mathrm{rs}} \sim {f_{\mathrm{exp}} E_{\mathrm{exp}}
\over 4\pi R_{\mathrm{s}}^3}\ ,
\label{eq:rspressure}
\end{equation}
where $f_{\mathrm{exp}}$ is the fraction of the supernova
explosion energy $E_{\mathrm{exp}}$ that is present as internal
energy of the gas between forward and reverse shock.
Plugging Eq.~(\ref{eq:rspressure}) into Eq.~(\ref{eq:rsradius1})
one obtains
\begin{eqnarray}
R_{\mathrm{rs}} & \sim & \sqrt{\,\left( 1-\frac{1}{\beta} \right) 
{\dot M_{\mathrm{w}} u_{\mathrm{w}} R_{\mathrm{s}}^3 \over
f_{\mathrm{exp}}E_{\mathrm{exp}}} }  \nonumber \\
& \approx &
4.14\times 10^3 \,\sqrt{{\dot M_{\mathrm{w},-5} u_{\mathrm{w},9}
R_{\mathrm{s},10}^3 \over f_{\mathrm{exp,0.1}}E_{\mathrm{exp,51}}}} 
\ \ [\mathrm{km}]\ .
\label{eq:rsradius2}
\end{eqnarray}
The numerical value was computed by taking $\beta = 7$ and
normalizing the shock radius to $10^{10}\,$cm, the explosion energy
to $10^{51}\,$erg, and the parameter $f_{\mathrm{exp}}$ to 0.1. 

Equations~(\ref{eq:entropy3}) and (\ref{eq:entropy4}), evaluated
with the numbers for the wind quantities from our numerical model,
describe the entropy jump at the reverse shock in the
simulations very well. The same is true for the reverse shock 
radius computed from Eq.~(\ref{eq:rsradius1}) with the pressure
behind the reverse shock, $P_{\mathrm{rs}}$, taken from the
hydrodynamic simulations. Equation~(\ref{eq:rsradius2}), however,
does not yield a satisfactory agreement and for some models fails
even qualitatively to reproduce the behavior of the reverse shock
radius as a function of time. The reason for this mismatch between
simulations and analytic approximation is mainly the fact that the
factor $f_{\mathrm{exp}}$ cannot be considered as a constant. Instead,
during the expansion of the supernova ejecta, $p{\mathrm{d}}V$ work
converts internal energy to kinetic energy of the matter swept up by the 
outgoing shock. Therefore, as time goes on, a smaller and smaller 
fraction of the explosion energy remains stored as internal energy 
in the layer between forward and reverse shock. As a consequence,
$f_{\mathrm{exp}}$ decreases during the simulated evolution. In the
first four seconds, $f_{\mathrm{exp}} \approx 1$ turns out to be a
good choice, but lateron $f_{\mathrm{exp}}$ drops monotonically to
$f_{\mathrm{exp}} \approx 0.25$ at ten seconds. Taking
this into account, Eq.~(\ref{eq:rsradius2}) also yields a good 
description of the reverse shock radius as a function of time.

\begin{figure*}[!tpb]
   \centering
   \begin{tabular}{lr}
     \includegraphics[width=7cm]{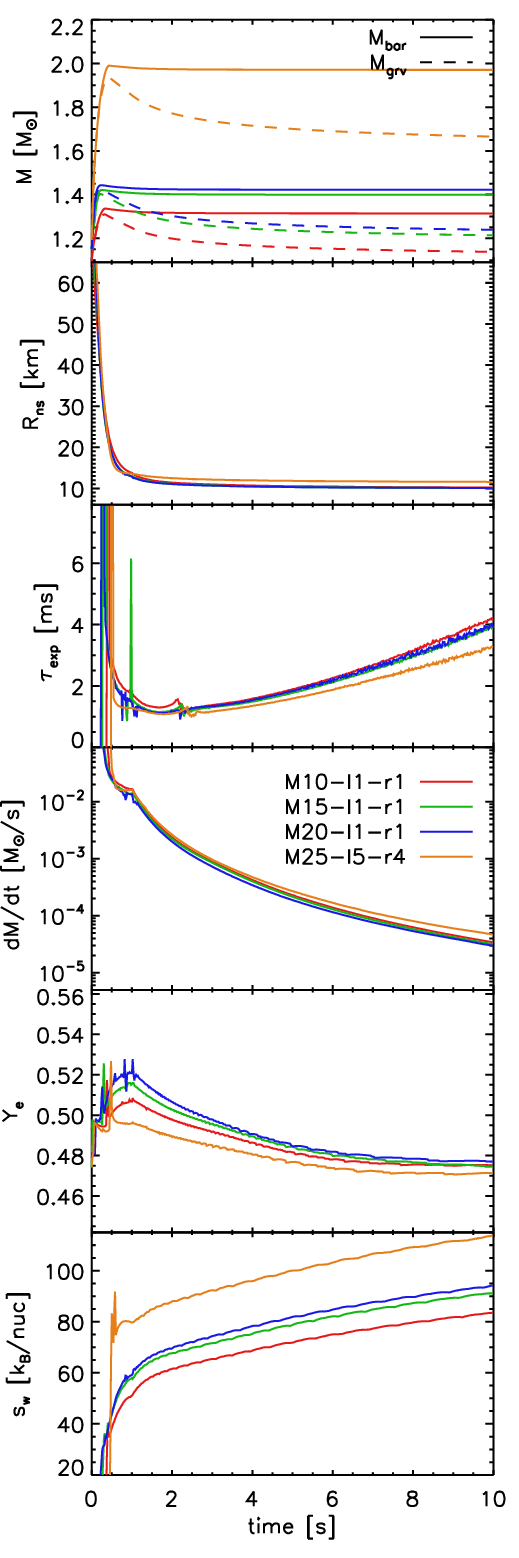} &
     \includegraphics[width=7cm]{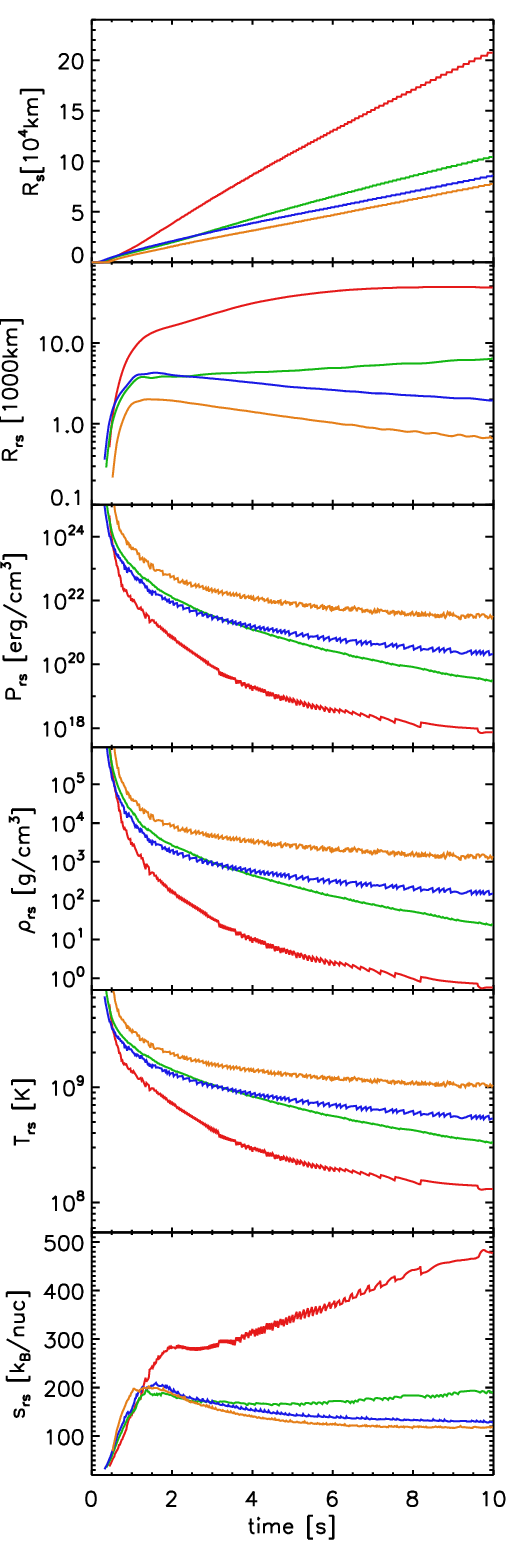}
   \end{tabular}
   \caption{Time evolution of different quantities for a
    set of simulations with different progenitor stars,
    M10-l1-r1, M15-l1-r1, M20-l1-r1, and M25-l5-r4. Shown
    are the baryonic mass, $M_{\mathrm{bar}}$,
    and gravitational mass, $M_{\mathrm{grv}}$ 
    (Eq.~\ref{eq:tovmassr}), neutron star radius,
    neutrino-wind expansion timescale according to 
    Eq.~(\ref{eq:tauthompson}), wind mass-loss rate, electron 
    fraction, and entropy per nucleon (left, from top to bottom), 
    radius of the supernova shock, radius of the reverse shock,
    and pressure, density, temperature, and entropy per nucleon 
    downstream of the reverse shock.}
   \label{fig:prog-timeevol}
\end{figure*}

\subsection{Different progenitors}
\label{sec:progenitors}

The analytic discussion of the previous section, in particular
Eqs.~(\ref{eq:entropy3}), (\ref{eq:entropy4}), and 
(\ref{eq:rsradius1}), allow us now
to understand the behavior of the wind termination shock in 
different progenitor stars. For this purpose we compare our
$15\,M_\odot$ reference model, M15-l1-r1, with models M10-l1-r1, 
M20-l1-r1, and M25-l5-r4, which are explosion simulations for
10.2, 20, and 25$\,M_\odot$ stars, respectively. The conditions
at the inner grid boundary were chosen such that the models have
similar explosion energies between roughly 1.3$\,$B and 2$\,$B
(Table~\ref{tab:finalresults}). 
The 25$\,M_\odot$ star has such a big mass accretion rate and 
correspondingly high accretion luminosity that the explosion
tends to become stronger than in the lower-mass progenitors.
To lessen this effect, we reduced the boundary luminosities 
compared to
the other models and chose a larger final radius of the inner 
boundary and thus of the new-born neutron star.

The neutron star mass and radius in the 10.2, 15, and 20$\,M_\odot$
simulations are rather similar (Table~\ref{tab:finalresults}) and so
are the time-dependent luminosities, mean energies, and energy 
radiated in $\nu_{\mathrm{e}}$ and $\bar\nu_{\mathrm{e}}$, 
as well as the total energy release in neutrinos of all
flavors, $\Delta E_{\mathrm{tot}}$, (Fig.~\ref{fig:prog-neutrinos}).
The $25\,M_\odot$ run, however, sticks out with significantly higher
values of all these quantities. Progenitor-dependent differences
associated with the density structure of the collapsing star outside
of the iron core are responsible for the differences in the
time-dependence of the explosion energy
seen between the 10.2, 15, and 20$\,M_\odot$ models in the lower
panel of Fig.~\ref{fig:prog-neutrinos}. A more massive progenitor
has a higher mass accretion rate and accretion luminosity and also
a larger mass in the gain layer. Its explosion therefore tends to
be more energetic. In case of the models M20-l1-r1 and
M25-l5-r4, the large binding energy of the outer stellar shells
later on leads to a visible decrease of the explosion energy from
a maximum value reached transiently during the simulation
(Fig.~\ref{fig:prog-neutrinos}).

\begin{figure}[tpb!]
  \centering
   \includegraphics[width=7.35cm]{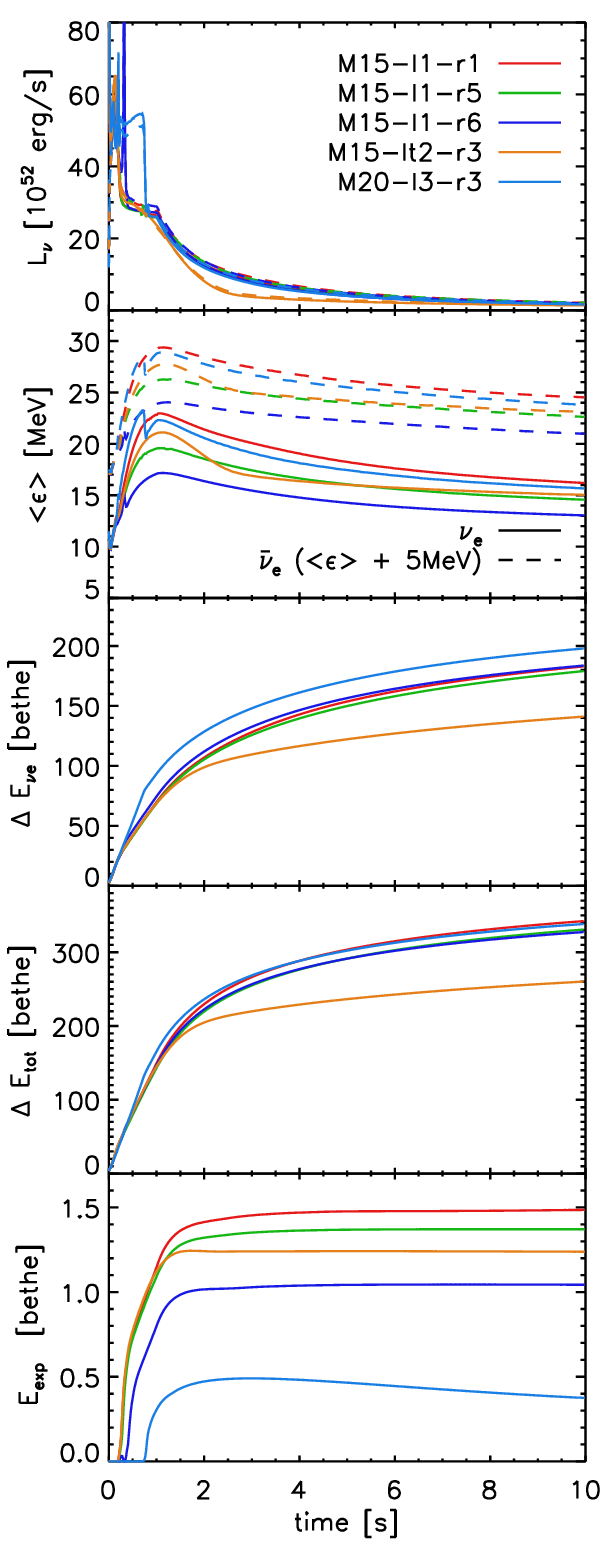}
  \caption{The same as Fig.~\ref{fig:prog-neutrinos}, but
    for simulations with varied inner boundary conditions
    (for clarity, however, we omitted the lines showing the
    rms energies of the radiated $\nu_{\mathrm{e}}$ and $\bar\nu_{\mathrm{e}}$
    and shifted the $\bar\nu_{\mathrm{e}}$ energies by 5$\,$MeV).
    The models M15-l1-r5, M15-l1-r6, M15-lt2-r3, and
    M20-l3-r3 are compared with our reference 15$\,M_\odot$
    model M15-l1-r1 in order to demonstrate the influence
    of different neutron star radii, neutron star masses,
    and core neutrino luminosities at the inner grid boundary.}
  \label{fig:boundvars-neutrinos}
\end{figure}

Figure~\ref{fig:prog-timeevol} shows the time evolution of quantities 
that determine and characterize the neutrino wind and reverse 
shock behavior in our simulations with different progenitors.
The wind properties (left column in Fig.~\ref{fig:prog-timeevol}) 
exhibit their well-known dependence on the 
neutron star mass and radius and on the neutrino luminosities and
mean energies. Because of the similarity of these quantities in 
case of the 10.2, 15, and 20$\,M_\odot$ models, only rather small 
differences are visible between these runs, revealing a slightly longer
expansion timescale, lower mass-loss rate, and higher entropy for
model M20-l1-r1 with its more massive neutron star 
(see also Table~\ref{tab:finalresults}). The electron fraction
shows a somewhat wider variation because of its strong sensitivity
to the spectral and flux differences of the $\nu_{\mathrm{e}}$ and 
$\bar\nu_{\mathrm{e}}$
emission. The larger neutron star mass and neutrino luminosities
in case of the 25$\,M_\odot$ progenitor separate this model clearly 
from the others. They affect in particular the neutrino-wind entropy,
which scales with the value of the neutron star mass but is only
weakly dependent on the neutrino emission properties (Eq.~\ref{eq:QWs}).
Nevertheless, since the neutron star is not extremely compact
($R_{\mathrm{ns}}\approx 11.5\,$km; Table~\ref{tab:finalresults})
and only moderately massive (gravitational mass $M_{\mathrm{grv}}\approx
1.66\,M_\odot$), the wind entropy is never higher than 
115$\,k_{\mathrm{B}}$ per nucleon during the 10 seconds of computed
postbounce evolution.
The expansion timescale and mass loss rate of the 25$\,M_\odot$
case are more similar
to the other models because of a partial cancellation of their 
dependences on $L$, $\epsilon$, and $M$ in Eqs.~(\ref{eq:QWtau})
and (\ref{eq:QWdotM}). 

The wind termination shock evolves largely differently in all 
cases (Fig.~\ref{fig:prog-timeevol}, right column). Obviously, the 
progenitor structure has a big influence on its behavior. The supernova
shock expands much faster in the lower-mass stars, causing a
more rapid decline of the pressure in the shell between forward and
reverse shock. The propagation of the forward
shock and the time-varying conditions there are communicated
inward to the reverse shock on the sound propagation
timescale. Therefore the pressure just downstream of the reverse
shock, $P_{\mathrm{rs}}$, as well as the density and temperature
at this location, decrease, too. From Eq.~(\ref{eq:rsradius1})
it can be understood that in model M10-l1-r1 the strong pressure 
reduction triggers a fast outward motion of the reverse shock. In the
15$\,M_\odot$ star the increase of $R_{\mathrm{rs}}$ is much less
extreme, and in the 20 and 25$\,M_\odot$ runs the wind termination 
shock even retreats after $\sim$2$\,$s of initial expansion and
transient stagnation. In these cases the decline of $P_{\mathrm{rs}}$
is not fast enough to compete with the decrease
of $\dot M_{\mathrm{w}}$ and $u_{\mathrm{w}}$ in the numerator
of Eq.~(\ref{eq:rsradius1}). A similar effect can be observed at
$t = 1\,$s when we change the time-dependence of the neutrino luminosity 
at the inner grid boundary. The subsequent luminosity decrease
leads to the mass-loss rate and velocity of the wind dropping
more quickly than $P_{\mathrm{rs}}$ in all models except M10-l1-r1,
explaining why the initial expansion of the reverse shock 
is stopped at about this time.

\begin{figure*}[!htb]
   \centering
   \begin{tabular}{lr}
     \includegraphics[width=7cm]{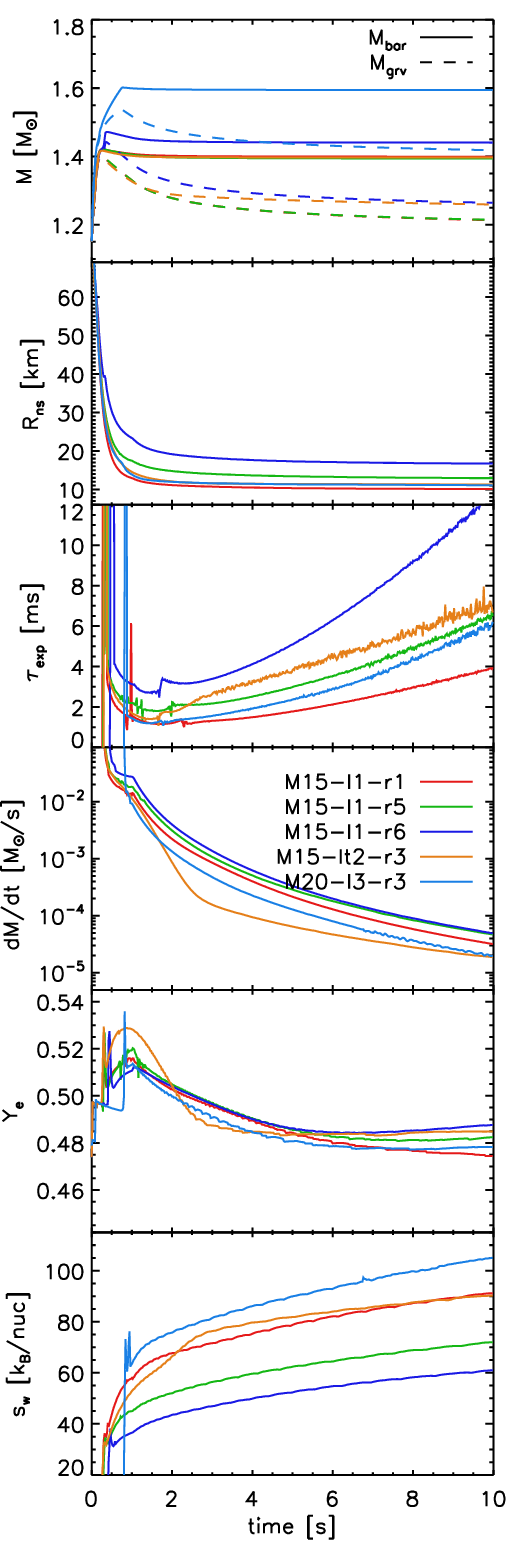} &
     \includegraphics[width=7cm]{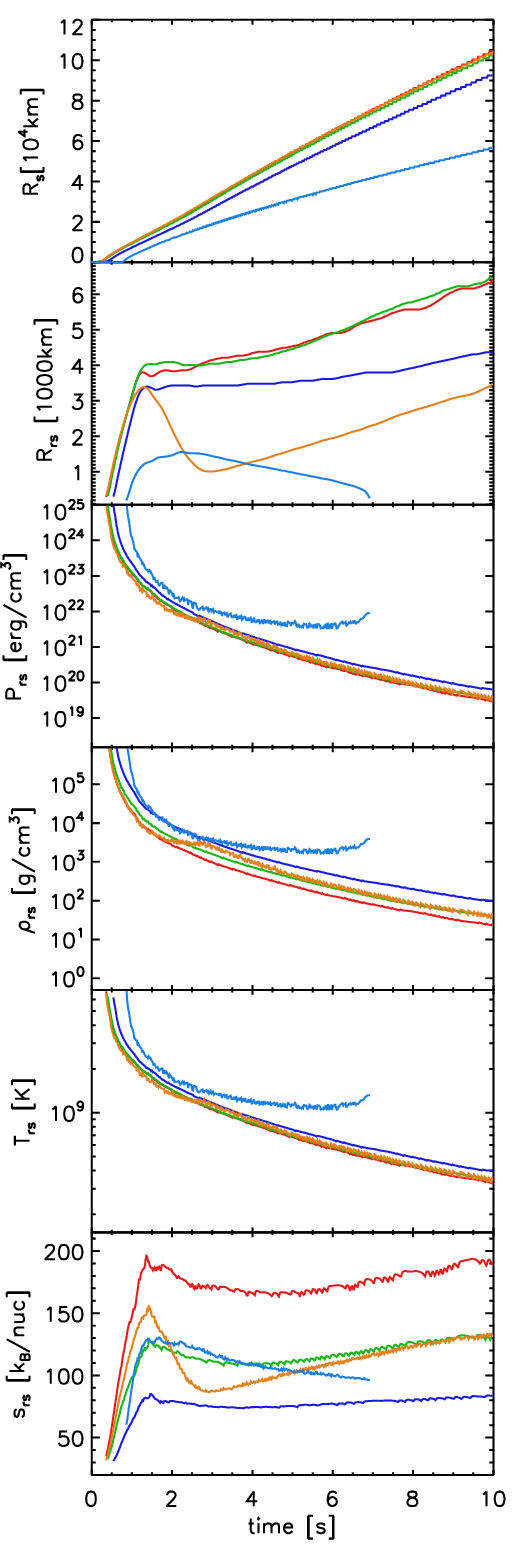}
   \end{tabular}
   \caption{The same as Fig.~\ref{fig:prog-timeevol}, but
    for simulations with varied inner boundary conditions.
    The models M15-l1-r5, M15-l1-r6, M15-lt2-r3, and
    M20-l3-r3 are compared with our reference 15$\,M_\odot$
    model M15-l1-r1 in order to demonstrate the influence
    of different neutron star radii, neutron star masses,
    and core neutrino luminosities at the inner grid boundary.}
   \label{fig:boundvars-timeevol}
\end{figure*}

Because of the different reverse shock behavior, the density,
temperature, and entropy downstream of the reverse shock 
as functions of time show also large differences between the 
progenitors (Fig.~\ref{fig:prog-timeevol}). 
In model M10-l1-r1 the wind termination
shock moves to radii beyond 10,000$\,$km within little
more than one second. During this phase the density 
$\rho_{\mathrm{w}}$ behind this shock drops to less than 
$10^3\,$g$\,$cm$^{-3}$ and the temperature $T_{\mathrm{w}}$ 
becomes lower than $10^9\,$K. The entropy, on the other hand,
is nearly 300$\,k_{\mathrm{B}}$ per nucleon after 2 seconds.
In the runs for the more massive progenitors, the density and 
temperature at the reverse shock are larger for a longer
period of postbounce evolution, and the entropy does not 
reach the very high values of the 10$\,M_\odot$ simulation.
The more massive the progenitor is --- or, more precisely,
the denser the shells around the iron core are --- the slower
propagates the shock for a given value of the explosion energy, 
and the more confined is the reverse shock. In none of the compared
cases, however, are the conditions at the wind termination shock 
constant with time.

\subsection{Variations of inner boundary conditions}
\label{sec:boundaryvariations}

It is clear from Eqs.~(\ref{eq:entropy3}), (\ref{eq:entropy4}), 
and (\ref{eq:rsradius1}) that the behavior of the reverse shock
does not only depend on the structure of the exploding star but
also on the neutrino-wind properties, in particular the wind 
mass-loss rate and velocity. Since the latter increases with the
distance from the neutron star, the radius of the reverse shock
introduces an additional velocity dependence in 
Eqs.~(\ref{eq:entropy3}), (\ref{eq:entropy4}), and (\ref{eq:rsradius1}).

In order to investigate the changes associated with different
strength and time evolution of the neutrino wind, we varied the
wind-determining parameters, i.e., the neutron star mass, radius,
and contraction, and the core neutrino luminosities and energies 
as functions of time. In this section we therefore discuss the
influence of these variations of the parameters used for the 
inner boundary condition.

The effect of the neutron star radius is visible from a comparison
of our reference 15$\,M_\odot$ model M15-l1-r1 with models M15-l1-r5 
and M15-l1-r6 in Fig.~\ref{fig:boundvars-timeevol}. These three 
simulations are computed with the same inner boundary condition for 
the neutrinos and produce neutron stars with approximately the same 
gravitational masses but final radii of 10, 13, and roughly 17$\,$km,
respectively (Table~\ref{tab:finalresults}). The neutrino luminosities
radiated from the nascent neutron star and the energy emitted in
$\nu_{\mathrm{e}}$ and $\bar\nu_{\mathrm{e}}$ as well as the total energy lost in neutrinos
are nearly the same (Fig.~\ref{fig:boundvars-neutrinos}). Because of
similar explosion energies, also the supernova shock in the three 
models propagates with similar velocity
(Fig.~\ref{fig:boundvars-timeevol}).

The mean neutrino energies, however, show a clear correlation with the neutron
star radius: the more compact the neutron star is, the higher are the energies
of the escaping $\nu_{\mathrm{e}}$ and $\bar\nu_{\mathrm{e}}$ in Fig.~\ref{fig:boundvars-neutrinos}.
Also the neutrino-wind properties reveal the variation with the compactness of
the neutron star that is qualitatively expected from the analytic expressions
given by \citet[][see also Sect.~\ref{sec:reference},
Eqs.~\ref{eq:QWs}--\ref{eq:QWdotM}]{Qian96}.  A larger $R_{\mathrm{ns}}$ leads
to a longer expansion timescale and thus lower wind velocity, larger mass-loss
rate, and smaller wind entropy (see Fig.~\ref{fig:boundvars-timeevol}). In case
of the mass-loss rate, however, the influence of the larger neutron star radius
is partly cancelled by the lower mean neutrino energies (see
Eq.~\ref{eq:QWdotM}) (and by the slightly higher neutron star mass of model
M15-l1-r6), for which reason the differences in $\dot M$ are rather modest, in
particular between models M15-l1-r5 and M15-l1-r6.

Qualitatively, the reverse shock exhibits the same behavior in these
two models as in M15-l1-r1. While its radius $R_{\mathrm{rs}}$
is essentially the same in models M15-l1-r1 and M15-l1-r5,
the wind termination shock, however, expands less strongly 
in model M15-l1-r6, reacting to the considerably lower wind velocity 
and slightly slower propagation of the supernova shock in this
somewhat less energetic model (cf.\ Eqs.~\ref{eq:rsradius1} and
\ref{eq:rsradius2}). Finally, the entropy of the 
matter decelerated in the reverse shock behaves as expected from
Eqs.~(\ref{eq:entropy3}) and (\ref{eq:entropy4}) when values for
the wind parameters
and reverse shock radius are inserted into these equations. 
It is highest in model M15-l1-r1
and lowest in model M15-l1-r6. The densities behind the reverse
shock are ordered inversely.

\begin{figure*}[htp!]
\sidecaption
\centering
\includegraphics[width=12cm]{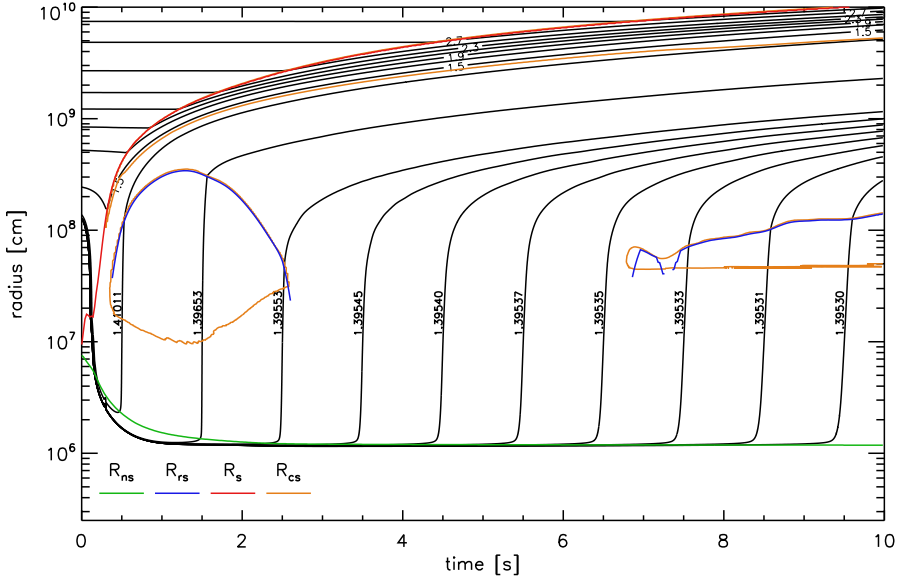}
  \caption[]{Same as Fig.~\ref{fig:M15-l1-r1:massshells}, but for
  model M15-lt1-r4. Here the explosion occurs at 0.22$\,$s after
  bounce and because of the assumed fast subsequent decay of the
  neutrino luminosity,
  the reverse shock reveals a much different behavior than in
  case of model M15-l1-r1. It temporarily disappears when a
  subsonic breeze instead of a wind develops after about 2.5$\,$s.
  At $t \ga 7\,$s the outflow expansion becomes supersonic again
  and a wind termination shock appears again.}
  \label{fig:M15-lt1-r4:massshells}
\end{figure*}

\begin{figure*}[!htb]
   \centering
   \begin{tabular}{lr}
     \includegraphics[width=7cm]{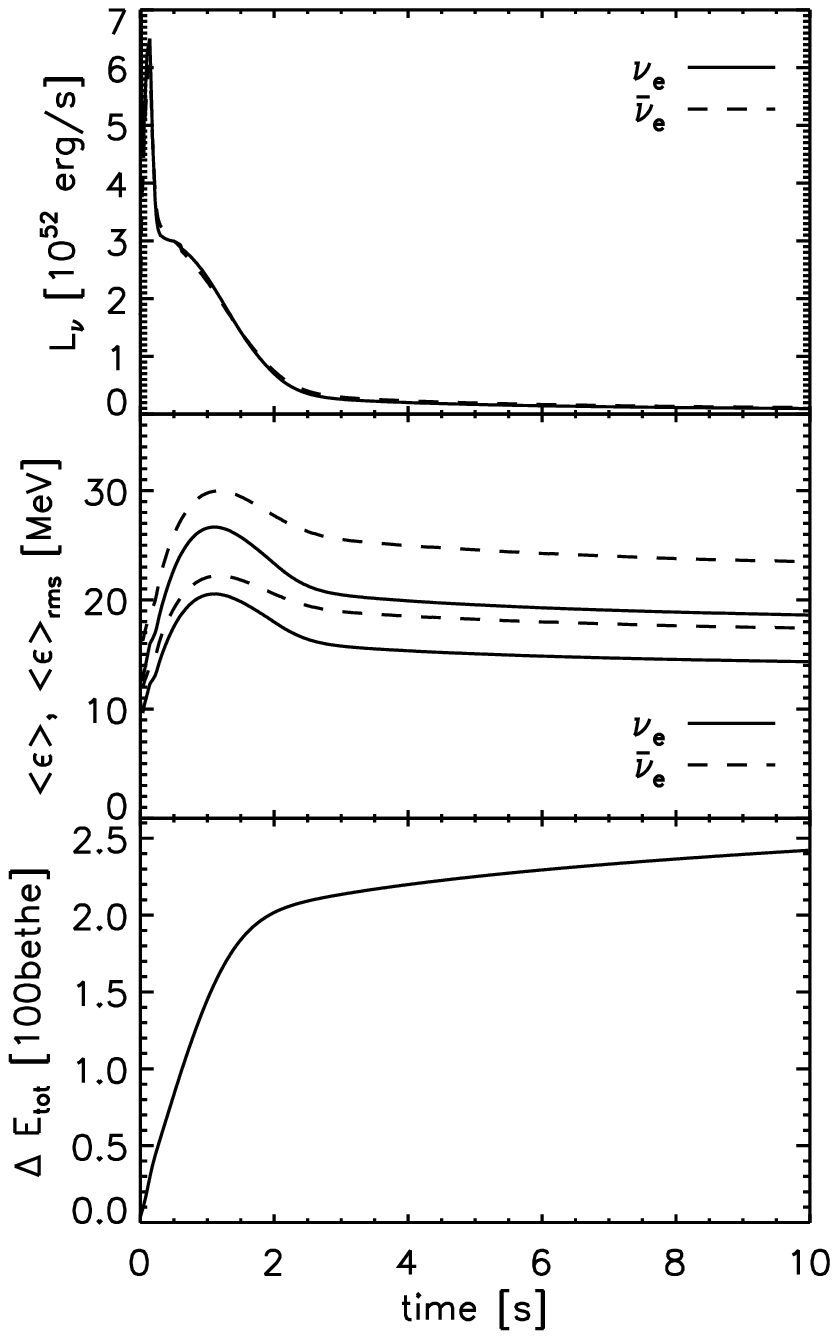} &
     \includegraphics[width=7cm]{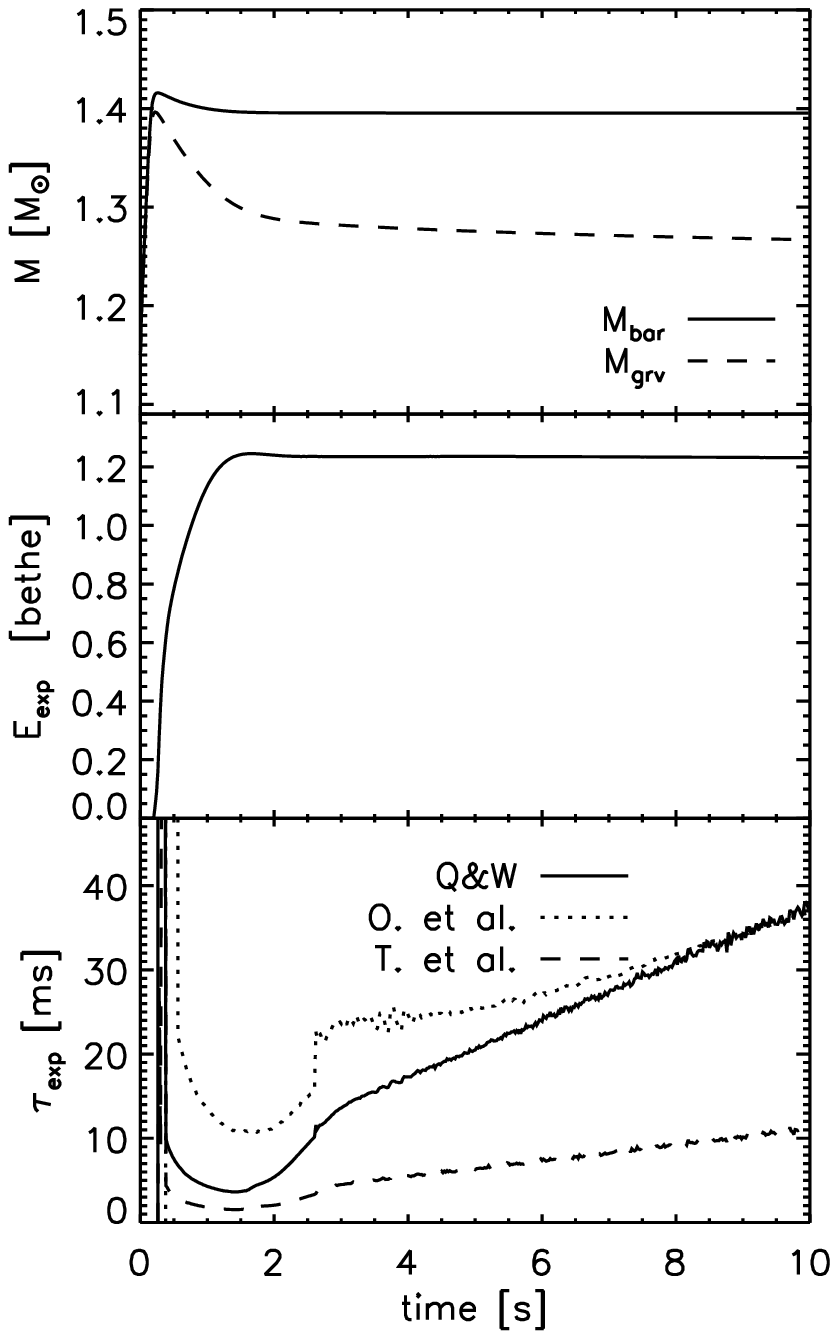}
   \end{tabular}
   \caption{Same as Fig.~\ref{fig:M15-l1-r1:neutrinos}, but for model
            M15-lt1-r4. Compared to model M15-l1-r1, the neutrino
            luminosities and mean energies decrease faster, the total
            energy radiated in neutrinos and the explosion energy are
            lower, and the gravitational mass of the neutron star is
            larger. The breeze solutions that develop between 2.5$\,$s
            and 7$\,$s have significantly longer expansion timescales.}
   \label{fig:M15-lt1-r4:neutrinos}
\end{figure*}

\begin{figure*}[!htb]
   \centering
   \begin{tabular}{lr}
     \includegraphics[width=7cm]{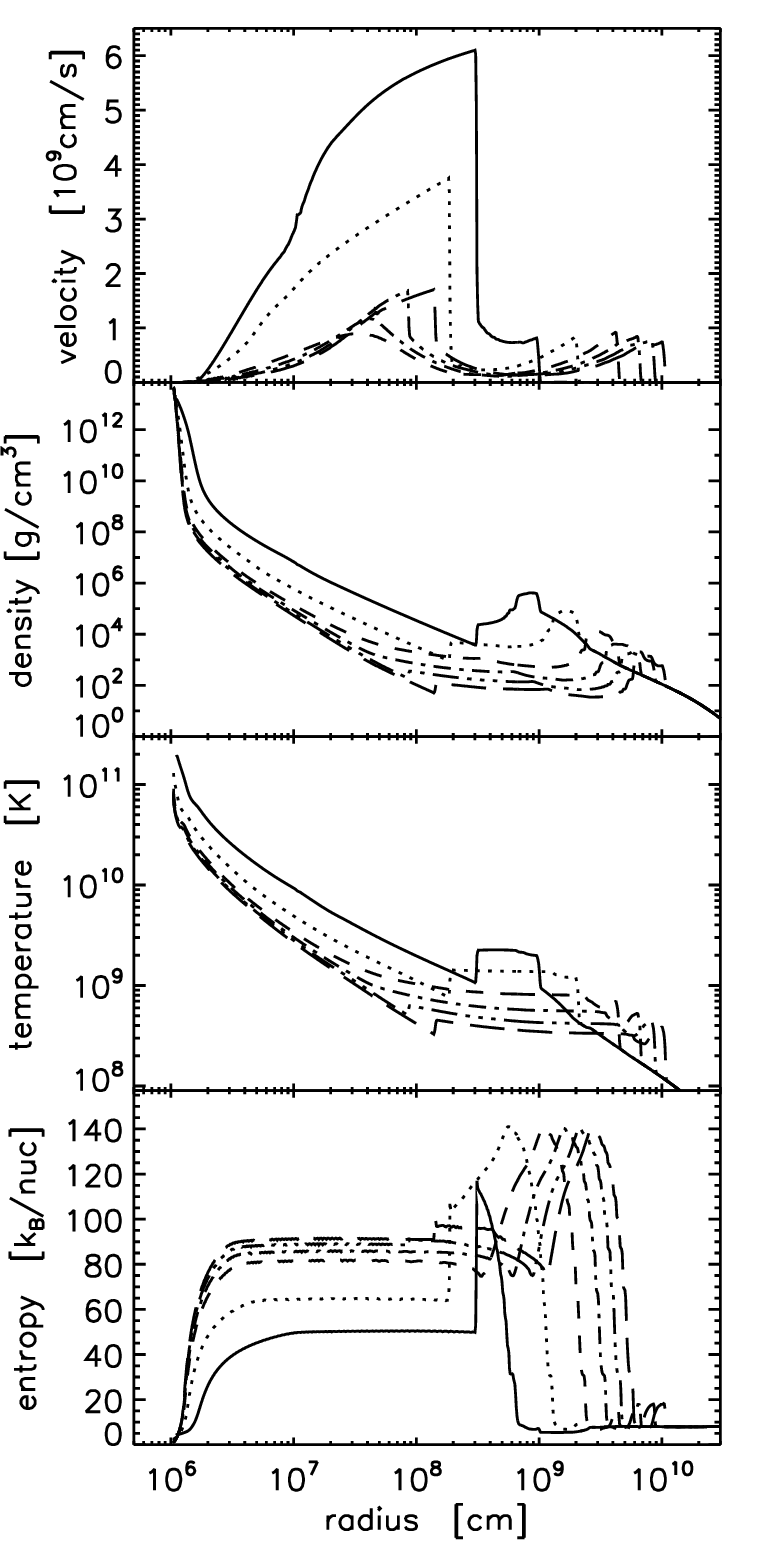} &
     \includegraphics[width=7cm]{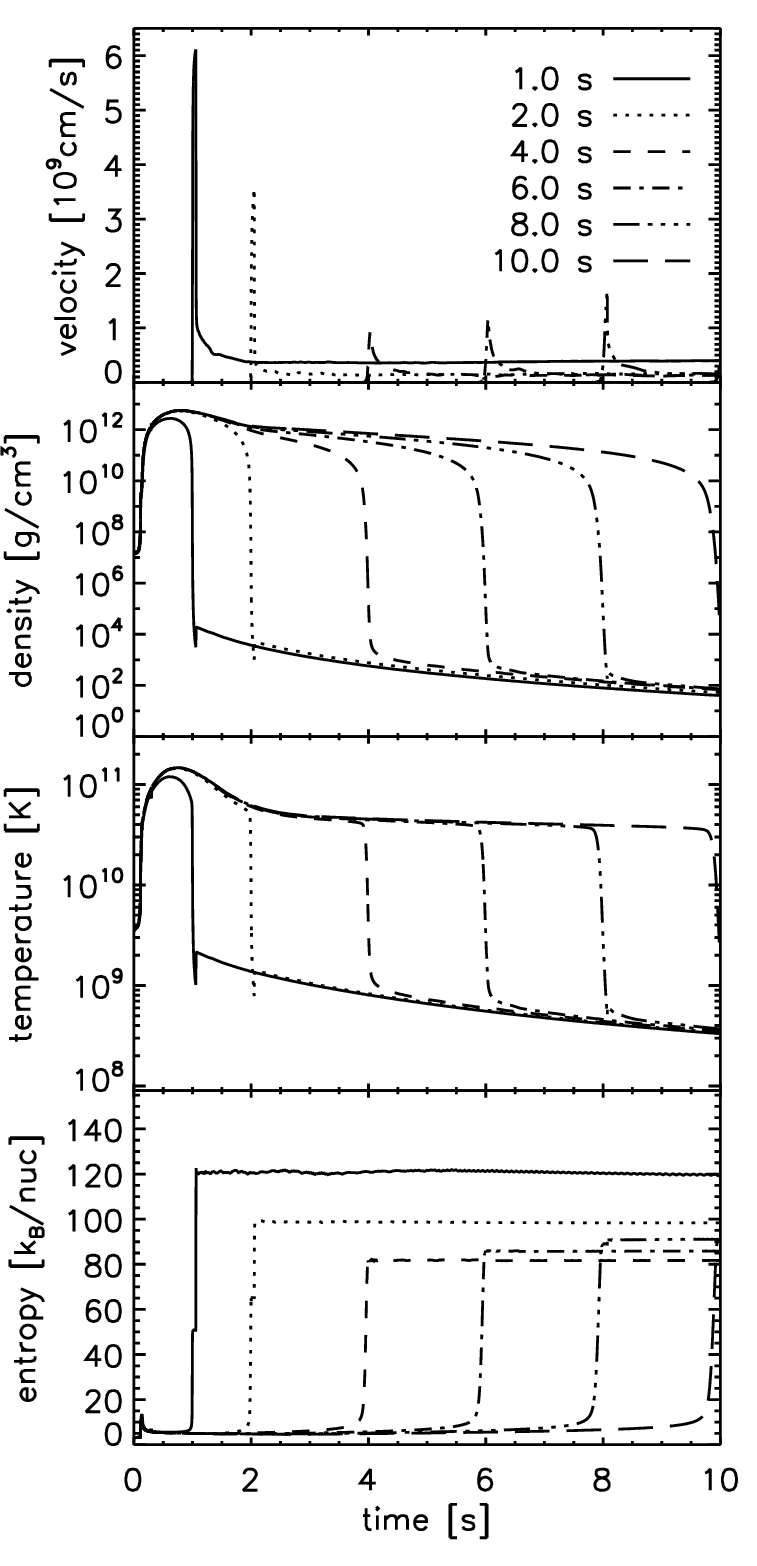}
   \end{tabular}
   \caption{Left, from top to bottom:
    Radial profiles of the velocity, density, temperature, and
    entropy in model M15-lt1-r4 at different postbounce times. The profiles
    should be compared with the corresponding ones of model M15-l1-r1 in
    Fig.~\ref{fig:M15-l1-r1:profiles}.
    Right: The same quantities as functions of time
    for some mass shells ejected in the neutrino-driven outflow of
    model M15-lt1-r4. For comparison with Model M15-l1-r1, see
    Fig.~\ref{fig:M15-l1-r1:trajectories}.}
   \label{fig:M15-lt1-r4:profsandtrajs}
\end{figure*}

Model M15-lt2-r3 demonstrates the influence of a more rapid decay
of the luminosities and mean energies of the radiated neutrinos
after one second of postbounce evolution.
This is associated with a reduced energy loss of the nascent neutron
star and leads to an increase of the wind expansion timescale, a 
steep decrease of the wind mass loss rate, and a higher wind entropy
compared to model M15-l1-r1. The reverse shock reacts to that by
a rapid recession between about 1$\,$s and 3$\,$s after bounce
(cf.\ Eq.~\ref{eq:rsradius1}) before it starts an outward motion
again at later times when the pressure $P_{\mathrm{rs}}$ drops
faster than $\dot M_{\mathrm{w}}$ and $u_{\mathrm{w}}$ of the wind.
Due to the small reverse shock radius, however, $u_{\mathrm{w}}$
at the shock is low and the entropy increase through the wind 
termination shock is modest. 

Model M20-l3-r3 with a neutron star radius and neutrino emission 
properties very similar to model M15-l1-r1, but a significantly 
higher neutron star mass, reveals an even more extreme behavior.
The larger neutron star mass increases the wind entropy, however
at the same time reduces the wind mass-loss rate and the inverse 
wind expansion timescale (and thus the wind velocity;
cf.\ Eqs.~\ref{eq:QWs}--\ref{eq:QWdotM}).
Moreover, the explosion energy of this model is very low
and the supernova shock expands only slowly. All together forces
the wind termination shock to retreat as the neutrino fluxes decay,
until it falls below the sonic point in the wind and disappears.
This brings the whole region from the proto-neutron star surface
to the outer boundary of the neutrino-driven outflow (which is the
contact discontinuity between shock-accelerated ejecta and 
neutrino-heated ejecta) in sonic contact, 
see Fig.~\ref{fig:M15-lt1-r4:massshells}. 
The neutrino-driven outflow is now only a subsonic breeze and
merges with the dense shell of ejecta behind the outgoing 
supernova shock without being accelerated to supersonic speed.

In order to study this phenomenon and its implications in more
detail, we triggered the occurrence of the breeze at a much
earlier time in model M15-lt1-r4, in which the neutrino luminosities
at the inner boundary were assumed to decay faster and the 
radiated neutrino energy is therefore even lower than in model
M15-lt2-r3 (compare Fig.~\ref{fig:M15-lt1-r4:neutrinos} 
with Fig.~\ref{fig:boundvars-neutrinos}). As a consequence,
the neutrino wind does not have sufficient power to keep the
wind termination shock at a large radius. In spite of a standard
explosion energy ($E_{\mathrm{exp}}\sim 1.2\,$bethe; 
Table~\ref{tab:finalresults} and Fig.~\ref{fig:M15-lt1-r4:neutrinos}) 
and fast propagation of the supernova shock, the
reverse shock begins to retreat already after 1$\,$s and
disappears after 2.5$\,$s (Fig.~\ref{fig:M15-lt1-r4:massshells})
whereas this happened only after 7$\,$s as in model M15-lt2-r3 
(Fig.~\ref{fig:boundvars-timeevol}).
In Fig.~\ref{fig:M15-lt1-r4:neutrinos} one sees that the 
transition to the subsonic breeze is accompanied by a considerable
growth of the expansion timescales calculated from 
Eqs.~(\ref{eq:tauqw}) and (\ref{eq:tauthompson}). The timescale
calculated from Eq.~(\ref{eq:tauotsuki}) exhibits even a sudden
increase which
occurs when the wind termination shock has retreated so much that
it is encompassed by the radial integral of Eq.~(\ref{eq:tauotsuki}).
The integral then includes shock-decelerated outflow which cools 
much more slowly. 

After about 7$\,$seconds, however, the sound speed in the 
neutron star surroundings has dropped and the dense ejecta shell
behind the supernova shock has moved outward sufficiently far
that the neutrino-driven outflow can again reach supersonic
velocities, despite much less powerful acceleration than in the
first two seconds after bounce (Fig.~\ref{fig:M15-lt1-r4:massshells}).
This is visible also in the radial profiles and mass shell
trajectories plotted in Fig.~\ref{fig:M15-lt1-r4:profsandtrajs},
where at late times ($t \ge 8\,$s) the discontinuity 
that characterizes the presence of a wind
termination shock appears again in all quantities. 
Because of the meanwhile low wind velocity
and very low mass-loss rate and therefore small reverse-shock
radius, this shock is at late times much weaker than it was in the
early phase. The associated density, temperature, and entropy
steps are correspondingly small
(Fig.~\ref{fig:M15-lt1-r4:profsandtrajs}).

During the breeze phase the outflow material is accelerated to a 
maximum velocity and then continuously decelerated again as it 
joins into the dense layer of ejecta behind the supernova shock. The 
mass-shell trajectories on the rhs.\ of 
Fig.~\ref{fig:M15-lt1-r4:profsandtrajs} illustrate this smooth
transition from the breeze expansion to the slower evolution
when the matter is added to the dense ejecta shell.

Models M15-lt1-r4 and M20-l3-r3 demonstrate clearly that the wind 
termination shock can be a transient feature and its presence
is very sensitive to the time-dependent conditions in the
neutrino-driven outflow and the expansion of the dense postshock
shell of supernova ejecta. Simulations with a consistent treatment
of the neutron star evolution and of the baryonic mass loss of the
nascent neutron star are needed to make definitive predictions of
the evolution of a given progenitor star. But even then such
predictions are handicapped by our incomplete knowledge of the 
high-density equation of state and of the corresponding
properties and neutrino emission of forming neutron stars.

\section{Possible implications for nucleosynthesis}
\label{sec:speculations}

Our work had the goal to investigate the fundamental aspects of the
neutrino-driven outflow from newly born neutron stars and of the
wind termination shock that forms by the interaction of the 
supersonic wind with the shell of slower moving
ejecta behind the supernova shock. We intended to explore the
dependence of this reverse shock on the
wind and the explosion properties in different progenitor stars. The
outflow conditions in our models were found to differ significantly
from the subsonic breeze solutions considered in many previous
nucleosynthesis studies. Detailed network calculations are needed
to analyse the consequences of these differences for the heavy-element
formation in the outflows. Although this is beyond the scope of our
present study, we want to add here a few speculative remarks about the
possible implications.

\subsection{Shocked winds vs. unshocked winds and breezes}
\label{sec:windandbreeze}

Two main differences of the outflows in our models are potentially 
relevant for the assembling of heavy nuclei, in particular the formation
of r-process elements in wind phases that develop a neutron excess.
Firstly, the winds accelerate to supersonic speeds and therefore
have shorter expansion timescales than breeze solutions, which by definition
remain subsonic everywhere. This affects also the temperature range from
7$\times 10^9\,$K down to 3$\times 10^9\,$K that is crucial for the formation
of seed nuclei from $\alpha$ particles and free nucleons through three-particle
reactions (triple alpha and $\alpha\alpha$n) and subsequent $\alpha$
captures. A faster expansion leads to less seed production and therefore an
increase of the neutron-to-seed ratio. This increase might be significant
for the modest values of
the wind entropy obtained at the end of our simulations (we found up to
100--120$\,k_{\mathrm B}$ per nucleon, see Table~\ref{tab:finalresults},
but 150$\,k_{\mathrm B}$ do not appear implausible at later times $t > 10\,$s
when the neutrino luminosities and mean energies have dropped further).
Note that \citet{Takahashi94}, for example, employed breeze solutions with
considerably longer expansion timescales in their nucleosynthesis studies.

Secondly, while the subsonic breezes gradually slow down after
they have reached their maximum velocity, the wind matter going through the
reverse shock is abruptly decelerated and its temperature, density, and
entropy are increased. In the shocked flow the density and temperature
then continue to drop on a much longer
timescale than during the rapid early expansion of the wind.
If the wind deceleration happens at conditions where free
neutrons are still present --- which is the case at temperatures around
$10^9\,$K in the breezes considered by \citet{Takahashi94} but might
be true at even lower temperatures in faster winds --- the changed
conditions may lead to a shift of the (n,$\gamma$)-($\gamma$,n) equilibrium
and thus of the r-process path during the later stages of r-processing.
This might alter the outcome of the nucleosynthesis compared to a 
freely expanding wind with its continuously growing velocity and 
compared to a gradually decelerated breeze. The results by
\citet{Takahashi94}, see e.g.\ Fig.~2 there, suggest that the strength of
the r-processing indeed depends on the dynamics of the outflow also
at relatively low temperatures.
In a most extreme situation (very rapid expansion, very high entropy) 
one might even imagine that neutrons remain unbound in the wind
because their capture timescale becomes longer than the expansion timescale
at some radius, whereas the shock-decelerated environment might give
all neutrons enough time to be absorbed by seed nuclei. 

Detailed network calculations are needed to clarify how strongly 
these differences really affect the heavy-element production
in high-entropy supernova outflows for the conditions found in
our models, and how the influence of these differences depends on 
the parameters that characterize the 
neutrino-driven wind and the reverse shock behavior. Such calculations
being currently unavailable, we are not able to draw definitive 
conclusions here. In particular it is not clear whether indeed, and if
so when, shock-decelerated supersonic winds can be 
more favorable for a strong r-process than previously studied breezes 
or unshocked winds. It is interesting, however, that \citet{Thompson01}
mentioned a strengthening of the production of third-peak r-process
nuclei due to a slower postshock expansion and a shock-increased 
temperature at the time the r-process freeze-out happens.

\subsection{Progenitor trends}
\label{sec:progtrends}

We have presented simulations for a number of progenitor stars and
have varied our prescription of the neutrino luminosities and of the
neutron star radius and contraction with time. All of our choices for
these conditions at the inner grid boundary
were constrained by our need to obtain
neutrino-driven explosions (and reasonable values for the
explosion parameters) in spherically symmetric models. This
required the assumption of sufficiently large boundary luminosities
or could be achieved by assuming a more compact neutron star. The latter
case leads to an enhanced release of gravitational binding 
energy and to increased
neutrino fluxes and thus stronger neutrino heating behind the shock.
The considered boundary settings were also motivated by the goal to 
demonstrate the different kinds of behavior that can occur for
the wind termination shock, depending on possible variations of
the properties of the neutrino-driven wind, of the explosion, 
and of the progenitor conditions. Identifying clearly the
influence of individual aspects, however, is not an easy task in 
simulations like these, where the different components that play a 
role are strongly coupled.

Conclusions on systematic variations with the stellar progenitor
are not only handicapped by
the limited set of investigated models, but in particular also by
the incomplete understanding of the supernova explosion mechanism and
of the neutron star equation of state. This
has the consequence that neither the explosion properties (energy,
ejecta velocity, explosion timescale and thus mass cut) nor the
properties of the compact remnant (e.g., its mass-radius relation
and release of gravitational binding energy) can be reliably predicted 
at the present time, implying that the neutrino emission
as a function of time (which may also be strongly influenced by 
convection inside the nascent neutron star) is not well known.
This in turn means
that not only the time evolution of the neutrino-driven wind is uncertain
(cf.\ Eqs.~\ref{eq:QWs}--\ref{eq:QWdotM}) but also the behavior and the
influence of the wind termination shock is not definitely determined
(due to the dependence
of Eqs.~\ref{eq:entropy4} and \ref{eq:rsradius1} on the explosion and 
wind parameters). Additional complexity comes from the fact that
multi-dimensional phenomena are important at least in some regions and 
during some phases of the evolution. Because of all
these uncertainties, we concentrated here on a
matter of principle study. Future simulations that include the neutrino
cooling of the nascent neutron star instead of making use of our inner
boundary condition will help to link and thus to reduce the involved degrees
of freedom. Nevertheless, they will still have to test different nuclear 
equations of state and will have to make assumptions about the onset and the 
energetics of the supernova explosion.

In order to assess the resulting uncertainties in the prediction
of the evolution of any particular progenitor model, we therefore
investigated in Sect.~\ref{sec:boundaryvariations} 
models where we varied the inner
boundary conditions within a reasonable (but maybe somewhat
``extreme'') range of possibilities. In spite of these  
systematic uncertainties, we think that
our ``standard'' (and preferred) set of models in 
Sect.~\ref{sec:progenitors} reveals some trends that either agree
with expectations or are likely to be confirmed by 
more complete and more detailed future studies.

On the one hand,
the neutron stars in more massive progenitors tend to be more
massive, simply because of the larger stellar cores and the
bigger mass infall rates after core bounce. This, of course, causes
more energy release in neutrinos and higher neutrino fluxes for
a longer period of time. Both the larger neutron star mass and the
differences in the neutrino emission affect the neutrino-driven
wind. In particular they lead to a tendency towards higher wind entropies
for more massive progenitor stars (but, astonishingly, no
significant differences in the wind expansion timescale and
mass-loss rate as functions of time; see Fig.~\ref{fig:prog-timeevol}).
On the other hand, more massive stars with their bigger and denser metal
cores cause a less rapid expansion of the supernova shock during the
computed phase of the explosion. The dense ejecta shell behind the
supernova shock then forces the wind termination shock to follow this
behavior (Fig.~\ref{fig:prog-timeevol}). 

The consequence is that in less massive
progenitors the reverse shock reaches large radii very quickly and
continues to travel outward. Although it can lead to a huge increase
of the entropy in the wind, because matter is decelerated by the termination 
shock from very large velocities, these effects happen after a short
period of postbounce evolution only at a point where the 
temperature and density are already so low
that the influence on the nucleosynthesis is probably marginal. In contrast, 
the reverse shock in more massive stars stays at a small radius during
the whole 10$\,$s of computed evolution and therefore for the most
massive progenitors the reverse shock is endangered 
to retreat and even collapse at a later stage of the
evolution when the neutrino-wind power has decreased. To see
this happening, we would have to run our ``standard'' simulations for
a significantly longer evolution period than just 10 seconds (with 
substantial demands on computer time because of time-step limitations and
the need to continuously increase the numerical resolution).
The fundamental possibility of a contraction of the reverse shock, however,
could be demonstrated by varying the boundary conditions in some of our
models such that the neutrino luminosities and neutrino-wind power
had dropped sufficiently already within the canonical 10$\,$s of
computed evolution (Sect.~\ref{sec:boundaryvariations}). 
The reverse shock staying at a small radius causes a more moderate
increase of the wind entropy but still decelerates the supersonic wind
abruptly and raises the density and temperature of the expanding matter. 
These discontinuous changes of the outflow conditions are time-dependent
and might have interesting but so far unexplored 
consequences for the nucleosynthesis.

We therefore conclude that the neutrino-driven outflows in our
models are supersonic winds in which the nucleosynthesis-relevant 
conditions are similar to the wind solutions investigated previously 
(e.g., by \citealt{Thompson01}). However, these winds are bounded at
some radius by the termination shock, which modifies the outflow
behavior drastically. In low-mass supernova progenitors 
this reverse shock is likely to have nucleosynthesis-relevant consequences
only during the first few seconds after bounce. 
For more massive progenitors the winds are abruptly decelerated, 
compressed, and heated by the wind termination
shock during a stage of their expansion (i.e. at a density and 
temperature) where the r-process path might be affected.
The combination of very rapidly expanding and ultimately supersonic 
wind and a termination shock that changes the evolution of the outflow,
has so far not been investigated in detail for its nucleosynthesis 
implications. In progenitors with very massive and dense metal cores
(roughly for stars with $\ga$25$\,M_\odot$), which can give birth to
neutron stars with a mass of $\ga$2$\,M_\odot$, the wind termination
shock might recede and possibly disappear so that the outflow 
develops into a subsonic breeze instead of being a supersonic wind.
In this case the nucleosynthesis environment becomes more similar
to the conditions studied by \citet{Takahashi94}, \citet{Sumiyoshi00}
or \citet{Otsuki00} where subsonic ejecta reach a maximum velocity
and then slow down gradually to meet an imposed boundary condition
at a certain radius. This boundary condition is supposed to account 
for the presence of the dense shell of ejecta that
follows the expanding supernova shock and that absorbs the later
neutrino-driven outflow from the cooling neutron star. Our models 
show that in a supernova core this ``boundary'' is time dependent
and its consequences are not likely to be well described by the simple 
(e.g. constant) assumptions made for the idealized neutrino-driven 
outflows considered in many nucleosynthesis studies. 

Our simulations
therefore suggest that an analysis of the of the nucleosynthesis based
on more detailed models of the conditions in supernova cores is very
desirable. If the reverse shock feature has a direct relevance for the
possibility of r-processing, its biggest impact must be expected for 
progenitors between roughly 15$\,M_\odot$ and roughly 25$\,M_\odot$,
where the wind termination shock is neither at too large radii nor 
disappears during the evolution phases that are most favorable for
a strong r-process.

\section{Summary and conclusions}
\label{sec:conclusions} 

We have presented long-time 1D hydrodynamic simulations of 
neutrino-driven 
explosions and post-explosion outflows for progenitors with
different masses. In these simulations the core of the 
shrinking nascent neutron star at neutrino optical depths 
larger than about 100 was replaced by an inner boundary whose
contraction was prescribed and where time-dependent neutrino 
fluxes were imposed such that the neutrino-energy deposition 
around the neutron
star produced explosions with a desired energy. The 
time-dependence of both the boundary motion and of the neutrino
luminosities was varied to investigate their influence on the 
high-entropy baryonic outflow and its interaction with the 
preceding supernova ejecta. Solving the Newtonian equations of
hydrodynamics, we included approximately the effects of 
relativistic gravity by employing the ``effective relativistic
potential'' of \citet{Marek06}. This approximation
yields very good agreement with fully relativistic calculations
during the postbounce accretion phase, and we found also nice
consistency of our approach with fully relativistic solutions
of stationary neutrino winds.

For the neutrino transport we performed a radial integration of the
frequency-integrated energy and lepton number equations, in which the neutrino
spectra were assumed to have a Fermi-Dirac non-equilibrium shape in the sense
that the neutrino spectral temperature could differ from the temperature of the
stellar medium \citep[see][]{Scheck06}. The neutrino luminosities and mean
energies are thus functions of time as well as radius in our simulations.

Because of the involved approximations and assumptions, our calculations
can only be suggestive but are not suitable for definitive predictions
of the nucleosynthesis-relevant conditions in dependence of the 
progenitor star. Our main goal was therefore a matter-of-principle
study of the interaction of the 
neutrino-driven baryonic outflow from the neutron star surface
with the slower dense shell of ejecta that is accelerated by the
outgoing supernova shock. The most important results are the following:
\begin{itemize}
\item
All of our models develop supersonic winds at least during some
phases of their evolution. These winds are bounded by a termination
shock in which the flow is abruptly decelerated and the entropy,
density, and temperature of the flow are strongly increased.
\item
The basic properties of this wind termination shock
can be understood from
simple analytic considerations using the shock-jump conditions
at this reverse shock. The entropy of the shock-decelerated matter 
increases with the wind velocity and is lower for high wind 
density. Therefore a large reverse shock radius is favorable
for a high entropy jump. The reverse shock radius increases 
with the mass-loss rate and velocity of the wind, but decreases
when the pressure behind the reverse shock is high. The latter
dependence links the behavior of the reverse shock to the 
propagation of the supernova shock and thus to the progenitor
structure and the explosion properties.
\item
The conditions at the reverse shock are progenitor-dependent
and usually strongly time-dependent and therefore the shock
effects are not well represented by an outer boundary condition 
with constant pressure \citep[e.g.,][]{Sumiyoshi00} or 
constant temperature \citep[e.g.,][]{Wanajo02}.
The conversion of kinetic energy to internal energy in the wind
termination shock can raise the entropy to several times the wind
entropy. We found the highest values (until 10$\,$s of postbounce
evolution) of nearly 
500$\,k_{\mathrm{B}}$ per nucleon behind the reverse 
shock (more than a factor of five increase) --- but also the 
lowest temperatures ($\la 10^9\,$K) and densities 
($\la 1000\,$g$\,$cm$^{-3}$) --- in case of the considered
10$\,M_\odot$ progenitor. In this star the supernova shock and the
reverse shock propagate outward very rapidly. In the considered
progenitors with masses of more than 15$\,M_\odot$ the maximum
entropies are higher than 200$\,k_{\mathrm{B}}$ per nucleon, 
corresponding to an increase of roughly a factor of three,
with densities and temperatures behind the reverse shock 
in the first ten seconds of
typically 100--$10^4\,$g$\,$cm$^{-3}$ and 0.4--$2\times 10^9\,$K,
respectively.
\item
When the supernova shock expands slowly (as in the case of very
massive progenitors with big and dense metal cores) 
or the neutrino emission from the nascent neutron star decays
rapidly and the wind power thus drops quickly, the reverse shock
can show phases of recession and can even fall below the sonic 
point in the wind. The outflow then becomes a subsonic breeze that 
merges smoothly with the ejecta shell behind the shock without any
jumps in the velocity and in the thermodynamic quantities. Changing
conditions around the neutron star can lead to a re-establishment
of a supersonic wind at later times.
\end{itemize}

The consequences of the deceleration of the neutrino-driven wind or breeze by
the dense shell behind the supernova shock were parametrized previously 
in many r-process studies by imposing a mostly time-independent
outer boundary condition in hydrodynamic
models (e.g., \citealt{Sumiyoshi00}, \citealt{Terasawa02}) or by selecting
those solutions of the steady-state equations that fulfill certain conditions
\citep[e.g.,][]{Otsuki00,Wanajo01,Wanajo02}. Here, however, we have
demonstrated that such prescriptions do in general not adequately account for
the effects that happen when the neutrino-driven outflow collides with the
preceding, slower postshock shell. A wind termination shock can not only alter
the wind entropy, density, and temperature by factors of a few, 
but also leads to a much slower expansion
of the shocked outflow after its deceleration. It will have to be explored in
detail by nucleosynthesis calculations how the combination of the very rapid
expansion of the supersonic winds and the changing conditions in the 
matter due to the reverse shock affect 
heavy-element formation in the high-entropy supernova ejecta.

All of our simulations exhibit the presence of proton-rich
conditions in the neutrino-driven outflow during the first
2--3 seconds of the explosion. Only later a neutron excess
develops in the wind matter. Our result of $Y_{\mathrm{e}} > 0.5$ in
the early neutrino-driven wind is 
qualitatively in agreement with models that employ Boltzmann 
neutrino transport \citep{Buras06a,Pruet05,Froehlich06}. 
We point out, however, that quantitatively meaningful 
calculations of the proton-to-neutron ratio require 
frequency-dependent
neutrino transport. Moreover, a reliable prediction of the 
turning point from proton excess to neutron excess and of the value
of $Y_{\mathrm{e}}$ at late postbounce times is not possible without
fully consistent cooling calculations of the proto-neutron star
instead of our inner grid boundary with prescribed time dependence 
of the neutrino fluxes. At best, our calculations can be indicative
for the trends which will also be found in such improved 1D models.

The varying conditions in the neutrino-driven wind and the
strong time- and progenitor-dependence of the behavior
of the wind termination shock and of its effects on the
wind raise a serious question: Do supernova cores provide
the robust environment for producing the extremely uniform
solar-system like r-process abundance pattern between the
Ba- and Pt-peaks observed in
ultra metal-poor stars \citep[see, e.g.,][]{Cowan06}? 
In addition to the variations that are present in
our models, multi-dimensional effects lead to long-lasting
anisotropic accretion and at the same time directed outflows 
(e.g., \citealt{Scheck06}). These introduce a stochastic
element in the supernova evolution during the first 
seconds of postbounce evolution 
\citep[see also][]{Burrows06a,Burrows06b}. The supernova ejecta
in different directions can develop largely different conditions
due to the strong anisotropy of the explosion mechanism and 
of the environment of the forming neutron star. Downdrafts and 
pockets of dense, low-entropy matter in the convective shell
behind the forward shock, for example, lag behind the overall 
outward expansion and cause a large-scale deformation of the
reverse shock that terminates the supersonic neutrino-driven
wind also in the multi-dimensional case. 
The varying radius and orientation of the
reverse shock relative to the radial direction lead to an
angular dependence of the properties of the shocked matter.
Therefore the amount of matter ejected with certain
conditions (e.g., entropy, expansion timescale after passing
the shock) differs between
spherically symmetric and multi-dimensional models. A detailed
hydrodynamic study in 2D is currently underway
\citep{Arcones07}. 

It is hard to see how this chaotic variability can allow for 
the robustness of environmental conditions needed for producing a
uniform abundance pattern of high-mass r-process elements
(even if some chunks of possible ejecta achieve to develop suitable 
conditions of high entropy and low $Y_{\mathrm{e}}$ as observed 
in the recent models of \citealt{Burrows06b}).
If supernovae are the main sources of the high-mass nuclei
beyond the $A\sim 130$ abundance peak
--- and a number of arguments have been 
made in support of that \citep[e.g.,][]{Cowan04} --- 
a solution of this puzzle may be that these nuclei are produced
in the later stages of the neutrino-driven wind, which are
unaffected by the turbulent initial phase of the explosion. 
The proton-richness of the early supernova ejecta
seen in our models in agreement with state-of-the-art simulations
with energy-dependent neutrino transport, and the transition
to n-richness at later times, yield support for
this argument. The spherically symmetric supersonic winds 
with their wind termination shocks simulated in this work may
be more characteristic of these late stages than of the early,
still turbulent phases of the explosion. Previous investigations, 
however, suggest that the conditions in (unshocked) neutrino-driven 
outflows are insufficient for a strong r-processing unless the
neutron star is very massive (around 2$\,M_\odot$) and very
compact ($\sim\,$9$\,$km; e.g., \citealt{Thompson01,Otsuki00}).

It is currently not clear whether the supersonic outflows in our
more ``realistic'' explosion models with the presence
of the reverse shock can ease this constraint and help 
establishing an r-process favorable environment for
less extreme assumptions of neutron star mass and radius.
Maybe this is realised only in a subset of progenitors where
the reverse shock is at a beneficial location. 
Such a requirement could single out rather discontinuously 
progenitors with particular core and explosion properties as 
favorable, while other progenitors do not develop suitable 
conditions for a strong r-processing. More simulations,
improved modeling, and detailed nucleosynthesis calculations
are needed to explore such possibilities.

\begin{acknowledgements}
We are grateful to S.~Woosley and A.~Heger for providing us with their
progenitor models, to A.~Marek for computing collapse and prompt
shock propagation phases with the \textsc{Vertex} neutrino-hydrodynamics
code, and to R.~Buras for assisting A.A.\ in implementing and testing
the effective relativistic potential in the code with inner grid 
boundary. We also thank K.~Langanke,
G.~Mart\'{\i}nez-Pinedo, and I.~Panov for stimulating discussions.
Support by the Sonderforschungsbereich 375 on
``Astro-Particle Physics'' of the Deutsche Forschungsgemeinschaft is
acknowledged. The computations were performed on the NEC SX-5/3C and
the IBM p690 ``Regatta'' of the Rechenzentrum Garching, and on
the IBM p690 ``Jump'' of the John von Neumann Institute for 
Computing in J\"ulich.
\end{acknowledgements}

{\small
\bibliographystyle{aa}
\bibliography{biblio}
}

\end{document}